\documentclass[twocolumn,epjc3]{svjour3} %
\smartqed  
\RequirePackage{graphicx}
\RequirePackage{mathptmx}      
\RequirePackage{latexsym}
\RequirePackage[nointegrals]{wasysym}
\RequirePackage{mathtools}
\RequirePackage{amsmath}
\RequirePackage{amssymb}
\RequirePackage{xcolor}
\journalname{Eur. Phys. J. C}
\begin{document}

\title{Prompt atmospheric neutrinos in the quark-gluon string model}

\author{
S. I. Sinegovsky\thanksref{e1,addr1,addr2}
  \and  
M. N. Sorokovikov\thanksref{e2,addr1,addr2} 
}

\thankstext{e1}{e-mail: sinegovsky@jinr.ru}
\thankstext{e2}{e-mail: sorokovikov@jinr.ru}


\institute{Joint Institute for Nuclear Research, Joliot-Curie, 6, Dubna, Moscow region, 141980, Russia \label{addr1}
         \and 
Irkutsk State University, Gagarin blv. 20, Irkutsk, 664003, Russia \label{addr2}
}

\date{Received: date / Accepted: date}

\maketitle

\begin{abstract} 
We calculate the  atmospheric flux of prompt neutrinos, produced in decays of the charmed particles at energies beyond 1 TeV.  Cross sections of the $D$ mesons and ${\rm\Lambda}^{+}_{c}$ baryons production in $pA$ and $\pi A$ collisions are calculated in the phenomenological quark-gluon string model (QGSM) which is updated using recent measurements of cross sections of the charmed meson production in the LHC experiments. 
A new estimate of the prompt atmospheric neutrino flux is obtained and compared with the limit from the  IceCube experiment, as well as with predictions of other charm production models.


\end{abstract}

\section{Introduction}
\label{intro}
At present time the operating neutrino telescopes focus on the detection of astrophysical high-energy  neutrino fluxes: IceCube, a cubic kilometer detector at the South Pole~\cite{IceCube16,IceCube17,IceCube18}, ANTARES~\cite{antares18,antares13} located in the Mediterranean Sea, and underwater Baikal Gigaton Volume Detector (Baikal-GVD), a cubic kilometer-scale array, which is currently under construction in lake Baikal~\cite{GVD18,GVD18a}.

The Baikal-GVD has a module structure and consists of functionally independent sub-arrays (clusters) of optical modules (OMs) and is designed to detect astrophysical neutrino fluxes at energies from a few TeV up to $100$ PeV. 
Five clusters have been already installed (every includes 288 OMs), current instrumented volume of the Baikal-GVD is the largest in the Northern Hemisphere ($\sim 0.25$ km$^3$), and the first high-energy neutrino induced events are reconstructed. 
The first phase (GVD-1) to be completed by 2021 and  will comprise 9 clusters (2592 OMs), the full-scale GVD with an instrumented volume about of $2$ km$^3$ will consist of $10^4$ light sensors.

The diffuse flux of high-energy astrophysical neutrinos was revealed in 2013 at IceCube detector~\cite{IceCube13a,IceCube14}, and for 6 years about 100 neutrino events were detected in IceCube experiment \cite{IceCube17}.
Another important discovery was made recently: on 22 September 2017 IceCube have detected the high-energy neutrino event coincident both in the direction and the  time with the gamma-ray flare from the blazar TXS 0506+056~\cite{IC-TXS0506+056}. The event was later confirmed~\cite{IceCube2} by the archival IceCube data which display an excess of high-energy neutrino events (against the atmospheric neutrino flux) between Sept. 2014 and Mar. 2015, that give $3.5\sigma$-evidence for neutrino flux from the direction of TXS 0506+056.  %
This supports the hypothesis that the blazar TXS 0506+056 is the  individual  source of high-energy neutrinos and, presumably, the source of high-energy cosmic rays. 

Essential progress has been achieved in experimental studies of astrophysical and atmospheric  neutrino fluxes, however the prompt atmospheric neutrinos have not yet been detected.
High-energy neutrinos arising from decays of mesons and baryons, produced in hadronic collisions of cosmic rays with Earth's atmosphere, compose the background against the neutrinos from  distant astrophysical sources. The atmospheric neutrinos comprise two components, which are distinguished by zenith-angle distributions and the energy spectra. The anisotropic component, arising from decays of pions and kaons, has the softer  spectrum (``conventional'' or $\pi,K$-neutrinos). The second component, quasi-isotropic flux  produced at higher energies, mainly in decays of short-lived heavy charmed mesons and baryons $D, {\rm\Lambda}^{+}_{c}$, is characterized by harder spectrum. This component (``prompt'' neutrinos) is most uncertain because of scarce measurements  and  wide spread in model predictions of  the charm production cross sections at very high-energies.  
\par  
 The high energy  interactions  of cosmic rays with the Earth's atmosphere are dominated by the soft processes with small momentum transfer, which are beyond the  scope of perturbative technique of the quantum chromodynamics (QCD). Perturbative QCD models of the charmed particle production encounter difficulties related to the nonperturbative dynamics contribution. 
Thus, the elaboration of phenomenological models beyond the pQCD is required for comprehensive research of the charm production in hadronic interactions at high energies.

The quark-gluon string model (QGSM)  was developed \cite{KaidalovTer-Martirosyan,KaidalovPiskunova_charm,Kaidalov2003} to describe the soft and semihard hadronic processes at high energies: it has been  applied for successful explanation of characteristics of mesons and baryons production in hadron-nucleon collisions. QGSM was one of the first models to estimate the  prompt atmospheric neutrino flux ~\cite{BNSZ}.   
The recent data on the cross sections of charmed particle production, obtained in experiments at the LHC~\cite{ALICE_Abelev,ALICE_Adam,ALICE_Acharya,ATLAS}, allow the QGSM free parameters updating.
The updated version of QGSM is applied to calculate the prompt neutrino flux in the neutrino energy range 1 TeV  -- 100 PeV.  The calculation is based on the hadronic cascade model~\cite{BNSZ,Vall,nss98} and cross sections of $D$ meson and ${\rm\Lambda}^{+}_{c}$ baryon production in $pA$- and $\pi A$-collisions which are computed with the updated QGSM. We compare our result with the constraint obtained in the IceCube 
experiment~\cite{IceCube16} as well as with predictions of the color dipole model (ERS)~\cite{ERS}, SIBYLL 2.3c~\cite{sib-2.3c_18}, the NLO pQCD models, BEJKRSS~\cite{BEJKRSS} and  GRRST~\cite{grrst}.

\section{Production of charmed particles in QGSM}   
\label{sec:QGSM}
 
The nonperturbative quark-gluon string model (QGSM) gives unified descriptions of the soft hadronic processes. The model is based on the reggeon calculus, the topological $1/N_c$-expan\-sion of the amplitudes and the color string dynamics (see for more details ~\cite{KaidalovTer-Martirosyan,KaidalovPiskunova_charm,KaidalovPiskunova_ZPhys86,Shabelski,Lykasov} 
and references therein). 
The QGSM, having a small number of parameters, succeeded in describing of the multiparticle  production in hadron-nucleus collisions at high energies~
\cite{KaidalovTer-Martirosyan,KaidalovPiskunova_charm,Kaidalov2003,KaidalovPiskunova_ZPhys86,Shabelski,Lykasov,Arakelyan95,Arakelyan,Kaidalov03_YaF}. 
\par
Inelastic processes in the QGSM are described by the reggeon exchange (planar diagrams) with the intercept     $\alpha_{R}(0)<1$ and by the supercritical pomeron exchange (cylinder-type diagrams) with the intercept $\alpha_{P}(0)=1+{\rm\Delta}$, where ${\rm\Delta}>0$. At high energies, the contribution of cylindrical diagrams dominates due to factor $(s/s_0)^{\rm\Delta}$ whereas  the contribution of planar diagrams decreases as $(s/s_0)^{\alpha_{R}(0)-1} \propto (s/s_0)^{-1/2}$ .

The  planar and cylinder-type diagrams involve quark-quark, quark-gluon and gluon-gluon scattering. The summation of  $qq$, $qg$ and $gg$-diagrams leads to the Regge-behavior of scattering amplitude. The $s$-channel cuttings of cylindric diagrams describe the multiparticle production, and in the $t$-channel, these diagrams correspond to gluon exchanges. It is argued~\cite{Kaidalov03_YaF} that cylindric diagrams lead to the pomeron pole as the color singlet made up of sea quarks and soft gluons. 
The pomeron can be related to a sum of the ladder diagrams with exchange of reggeised gluons.
Sum of gluon ladders with possible quark loop insertions may produce the pomeron trajectory. 
The simplest two-gluons exchange leads to pomeron with the intercept $\alpha_{P}(0)=2S_g-1=1$.
 Account of mixing between $gg$ and $q\bar q$ Regge trajectories (glueballs and $q\bar q$ resonances) at the small $t$,  and effects of the small distance perturbative dynamics lead to the supercritical pomeron with  $\rm\Delta=\alpha_{P}(0)-1 \sim 0.15-0.25$, which ensures the Froissart behaviour of the total cross section.

To calculate inclusive cross sections of charmed hadron production, one needs know the distribution functions of the quarks of the colliding particles and the fragmentation functions of the quarks and diquarks. 
The inclusive cross sections of charmed hadron production are defined as the convolution of distribution functions of the valence (and the sea) quarks and diquarks of the colliding particles with the functions of quarks (diquarks) fragmentation into a charmed hadron. These functions are expressed in terms of intercept $\alpha_{R}(0)$ of the Regge trajectory ($\alpha_{R}(t)\simeq\alpha_{R}(0)+\alpha^{\prime}_{R}t$ in linear approximation), including the $\alpha_{\psi}(t)$ trajectory of the $c\bar{c}$ bound states. 
The complete set of distributions  and fragmentation functions can be found  in  Refs.~\cite{KaidalovPiskunova_charm,KaidalovPiskunova_ZPhys86,Shabelski,Lykasov,Arakelyan95}.
\par

For a nucleon target, the inclusive cross section of production of a hadron 
$h$ ($h=D^+$,\, $D^-$,\, $D^0$,\, ${\bar D}^0$,\, ${\rm\Lambda}^{+}_c$) 
can be written as a sum over $n$-pomeron cylinder  diagrams: 
\begin{equation}
\label{equation:inclusiveCS}
\tilde x\frac{d\sigma}{dx}=\int E\frac{d^3\sigma}{d^3p}d^2p_{\perp}=\sum\limits_{n=0}^{\infty}\sigma_{n}(s)\varphi_{n}^{h}(s,x),
\end{equation}
\noindent
where  $\sigma_n(s)$ is  the  cross  section  of the $2n$-strings (chains) production, corresponding   to  the  $s$-channel discontinuity of the multipomeron diagrams ($n$ cut pomerons and arbitrary number of external pomerons taking part in the elastic rescattering);  
$\varphi_{n}^{h}(s,x)$ is the $x$-distribution of the hadron $h$ produced  in  the fission  of $2n$ quark-gluon  strings: $\varphi_{0}^{h}(s,x)$ accounts of the contribution of the diffraction  dissociation of colliding  hadrons, $n=1$ corresponds to the strings formed  by valence  quarks  and diquarks, terms with  $n > 1$  are  related to  sea quarks  and antiquarks;
$x=2p_{\parallel}/\sqrt{s}$ is the Feynman variable, $p_{\parallel}$ is the longitudinal momentum of the produced hadron, $\sqrt{s}$ is the total energy of the two colliding hadrons in the c.m.f., \,  
$\tilde x =\sqrt{x_\bot^{2}+x^2}$,
 $x_{\perp}=2m_\bot/\sqrt{s}$,\, $m_\bot=\sqrt{<p_\bot^{2}>+m_{h}^{2}}$, \, $<p_\bot^{2}>$ is the mean square of the transverse momentum, and $m_{h}$ is the mass of the hadron $h$.  
\par
The cross sections $\sigma_{n}(s)$ were calculated ~\cite{Ter-Martirosyan} in the quasieikonal approximation which accounts of the low-mass diffractive excitations of the colliding particles and corresponds 
to maximum inelastic diffraction consistent with the unitarity condition. Only nonenhanced graphs were considered with neglect of interactions between pomerons: 
\begin{equation}
\sigma_{n}(s) = 
\frac{\sigma_{P}(s)}{nz(s)}\Bigg[1-e^{-z(s)}\sum\limits_{k=0}^{n-1}\frac{[z(s)]^k}{k!}\Bigg], \, \, (n\geq 1), 
\end{equation}
where  
\begin{align}
\sigma_P(s) = 8\pi\gamma_{P}(s/s_0)^{\rm\Delta} &, & 
z(s)=\frac{2C\gamma_{P}(s/s_0)^{\rm\Delta}}{R^2+\alpha_{P}^{\prime}\ln(s/s_0)}. 
\end{align}
\noindent 
Here $\sigma_P(s)$ is the pomeron contribution to the total cross section, $z (s)$ is the function representing a relative contribution of the successive rescatterings, $\gamma_{P}$ and $R^2$ are characteristics of the pomeron residue; $s_0=1 \; {\rm GeV}^2$.
The term with $n=0$ in \eqref{equation:inclusiveCS} corresponds to the elastic scattering and the diffractive dissociation: 
$\sigma_{0}(s)=\sigma_{\rm el}+\sigma_{\rm DD}=\sigma_{P}(s)[f(z/2)-f(z)],$
where  
\begin{equation} 
f(z(s))=\frac{1}{z(s)}\int\limits_{0}^{z(s)}\frac{1-e^{-y}}{y}dy.
\end{equation}
\noindent
The parameter $C=1+\sigma_{\rm DD}/\sigma_{\rm el}>1 $ takes into account the
the low-mass diffractive dissociation. 

The values of above parameters are found from experimental data on the total and differential cross sections of elastic $pp$ and  $p\bar p$ scattering at high energies~\cite{Kaidalov2003,Shabelski,Kaidalov03_YaF,Bleibel16}:  
\begin{gather*} 
\gamma_{P}^{pp}=1.27 \ {\rm GeV}^{-2}, \; R_{pp}^{2}=4.0 \ {\rm GeV}^{-2}, \; C_{pp}=1.8, \\
\gamma_{P}^{\pi p}=1.07 \ {\rm GeV}^{-2}, \; R_{\pi p}^{2}=2.48 \ {\rm GeV}^{-2}, \; C_{\pi p}=1.65, \\
{\rm\Delta}=0.156, \; \alpha^{\prime}_{P}=0.25 \ {\rm GeV}^{-2}.
\end{gather*}

In the case of $D$ meson production in $pp$ interaction, the functions $\varphi_{n}^{h}(s,x)$  can be written~\cite{KaidalovPiskunova_charm} as follows:
\begin{multline}
\label{equation:phip}
\varphi_{n}^{D}(s,x)= a^D\Big\{F_{q_V}^{D(n)}(x_{+})F_{qq}^{D(n)}(x_{-}) + F_{qq}^{D(n)}(x_{+})F_{q_V}^{D(n)}(x_{-})  \\  + 2(n-1)F_{q_{\rm sea}}^{D(n)}(x_{+})F_{q_{\rm sea}}^{D(n)}(x_{-})\Big\},
\end{multline}
where $x_{\pm}(s)=\frac{1}{2}\left[\sqrt{x^2+4m_{\perp}^{2}/s} \pm x\right]$.

For $\pi^{-}p$ interactions~\cite{KaidalovPiskunova_charm,Shabelski}: 
\begin{multline}
\label{equation:phipi}
\varphi_{n}^{D}(s,x)=a^D\Big\{F_{\bar q_{V}}^{D(n)}(x_{+})F_{q_{V}}^{D(n)}(x_{-}) +  F_{q_{V}}^{D(n)}(x_{+})F_{qq}^{D(n)}(x_{-}) \\ + 2(n-1)F_{q_{\rm sea}}^{D(n)}(x_{+})F_{q_{\rm sea}}^{D(n)}(x_{-})\Big\}.
\end{multline}
\noindent
The functions $F_{q_{V}}^{D(n)}(x_\pm)$, $F_{\bar{q}_{V}}^{D(n)}(x_\pm)$, $F_{qq}^{D(n)}(x_\pm)$, and $F_{q_{\rm sea}}^{D(n)}(x_\pm)$ defined as convolution of the quark distributions with the fragmentation functions, take into account contributions of the valence quarks, diquarks, and sea quarks. For example, in $pp$ collisions~\cite{KaidalovPiskunova_ZPhys86,Shabelski,Lykasov}:
\begin{multline}
F_{q_{V}}^{D(n)}(x_\pm) = \frac{2}{3}\int\limits_{x_\pm}^{1} f_{p}^{u_{V}(n)}(x_1) G_{u}^{D}(x_\pm/x_1) dx_1 + \\ + \frac{1}{3}\int\limits_{x_\pm}^{1} f_{p}^{d_{V}(n)}(x_1) G_{d}^{D}(x_\pm/x_1) dx_1,
\end{multline}
\begin{multline}
F_{qq}^{D(n)}(x_\pm) = \frac{2}{3}\int\limits_{x_\pm}^{1} f_{p}^{ud(n)}(x_1) G_{ud}^{D}(x_\pm/x_1) dx_1 + \\ + \frac{1}{3}\int\limits_{x_\pm}^{1} f_{p}^{uu(n)}(x_1) G_{uu}^{D}(x_\pm/x_1) dx_1.
\end{multline}
In case of  $\pi^{-} p$ interactions~\cite{Shabelski,Lykasov}:
\begin{equation}
F_{q_{V}}^{D(n)}(x_+) = \int\limits_{x_+}^{1} f_{\pi}^{d_{V}(n)}(x_1) G_{d}^{D}(x_+/x_1) dx_1,
\end{equation}
\begin{equation}
F_{\bar{q}_{V}}^{D(n)}(x_+) = \int\limits_{x_+}^{1} 
f_{\pi}^{\bar{u}_{V}(n)}(x_1) G_{\bar{u}}^{D}(x_+/x_1) dx_1,
\end{equation}
\noindent
where $f_{p}^{j}(x)$, $f_{\pi}^{j}(x)$ are the distribution functions of quarks, antiquarks and diquarks in colliding hadrons, $j=q,\bar{q},qq$; $G_{j}^{D}(x/x_1)$ are the fragmentation functions. 
At limits $x\rightarrow 0$ and $x\rightarrow 1$ these functions are defined by Regge asymptotics, and for the intermediate values of $x$ the interpolation is used~\cite{KaidalovPiskunova_charm,KaidalovPiskunova_ZPhys86,Shabelski}. In particular, 
\begin{equation}
f_{p}^{u_{V}(n)}(x) = C^{u_{V}}_n x^{-\alpha_{R}(0)} (1-x)^{\alpha_{R}(0)-2\alpha_{N}(0)+n-1},
\end{equation}
\noindent
\begin{equation}
G_{d}^{D^{-}}(x/x_1) = G_{\bar{u}}^{D^{0}}(x/x_1) = 
(1-x/x_1)^{\lambda-\alpha_{\psi}(0)}[1+a_1(x/x_1)^2],
\end{equation}
\noindent
where \, $\alpha_{R}(0)=0.5$, \, $\alpha_{N}(0)=-0.5$, \, $\alpha_{\psi}(0)=-2.2$, \ \, $\lambda=2<p_{\perp}^{2}>\alpha^{\prime}_{R}=0.5$, and the coefficient $C^{u_{V}}_n$ is determined by normalization $\int\limits_{0}^{1} f_{p}^{u_{V}(n)}(x) dx=1$. More details on the functions $\varphi_{n}^{h}(s,x)$, $f_{p}^{j}(x)$ and $G_{j}^{D}(x/x_1)$ can be found in~\cite{KaidalovPiskunova_charm,KaidalovPiskunova_ZPhys86,Shabelski,Lykasov,Arakelyan95}.
\par
The distribution function $\varphi_{n}^{h}(s,x)$ for the case of ${\rm\Lambda}_c$ production in $pp$ collisions can be written~\cite{KaidalovPiskunova_ZPhys86,Piskunova_Lambda} as following:
\begin{multline}\label{equation:phipLc1}
\varphi_{n}^{{\rm\Lambda}_c}(s,x) = a_1^{{\rm\Lambda}_{c}} [F_{1qq}^{{\rm\Lambda}_c(n)}(x_{+}) + F_{1qq}^{{\rm\Lambda}_c(n)}(x_{-}) ] + \\ + a_0^{{\bar{\rm\Lambda}}_{c}} [F_{q}^{{\rm\Lambda}_c(n)}(x_{+}) F_{0qq}^{{\rm\Lambda}_c(n)}(x_{-}) + F_{0qq}^{{\rm\Lambda}_c(n)}(x_{+})F_{q}^{{\rm\Lambda}_c(n)}(x_{-}) + \\ + 2(n-1) F_{q_{\rm sea}}^{{\rm\Lambda}_c(n)}(x_{+}) F_{\bar{q}_{\rm sea}}^{{\rm\Lambda}_c(n)}(x_{-})]. 
\end{multline}
Here $F_{1qq}$ denotes the distribution at the leading diquark fragmentation with weight $a_1^{{\rm\Lambda}_{c}}$ and $F_{0qq}$ is the distribution for the nonleading fragmentation of  diquarks  written with the central density parameter  $a_0^{{\bar{\rm\Lambda}}_{c}}$: 
\begin{multline}
F_{1qq}^{{\rm\Lambda}_c(n)}(x_{\pm}) = \frac{2}{3}\int\limits_{x_{\pm}}^{1} f_{p}^{ud(n)}(x_1) G_{1ud}^{{\rm\Lambda}_c}(x_{\pm}/x_1) dx_1 + \\ + \frac{1}{3}\int\limits_{x_{\pm}}^{1} f_{p}^{uu(n)}(x_1) G_{1uu}^{{\rm\Lambda}_c}(x_{\pm}/x_1) dx_1,
\end{multline}
\begin{figure} [!b] 
	\centering
	\includegraphics[trim=0cm 0cm 0cm 0cm, width=0.46\textwidth]{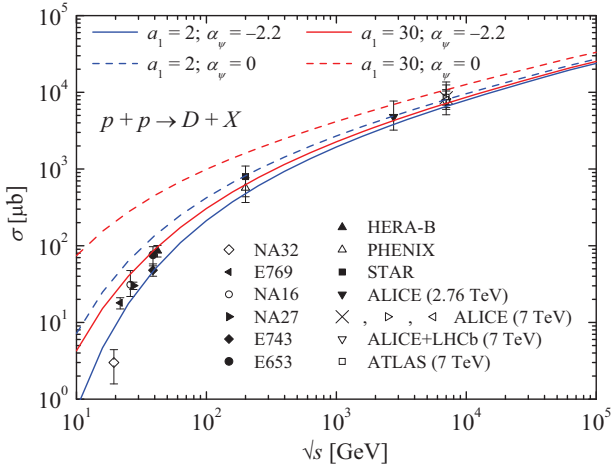}     
	\caption{\label{fig:1} Total cross sections of~$D/\bar{D}$ mesons production in \textit{pp} collisions. QGSM calculations: $\alpha_{\psi}(0)=-2.2$ (solid lines), $\alpha_{\psi}(0)=0$ (dashed lines); blue (bottom): $a_{1}=2$;  red: $a_{1}=30$. The experimental data are taken from Refs. \cite{ALICE_Abelev,ALICE_Adam,ALICE_Acharya,ATLAS,NA32,E769,NA16,NA27,E743,E653,HERA-B,PHENIX,STAR}.} 
\end{figure}
\begin{multline}
F_{0qq}^{{\rm\Lambda}_c(n)}(x_{\pm}) = \frac{2}{3}\int\limits_{x_{\pm}}^{1} f_{p}^{ud(n)}(x_1) G_{0ud}^{{\rm\Lambda}_c}(x_{\pm}/x_1) dx_1 + \\ + \frac{1}{3}\int\limits_{x_{\pm}}^{1} f_{p}^{uu(n)}(x_1) G_{0uu}^{{\rm\Lambda}_c}(x_{\pm}/x_1) dx_1,
\end{multline}
\begin{multline}
F_{q}^{{\rm\Lambda}_c(n)}(x_{\pm}) = \frac{2}{3}\int\limits_{x_{\pm}}^{1} f_{p}^{u_{V}(n)}(x_1) G_{u}^{{\rm\Lambda}_c}(x_{\pm}/x_1) dx_1 + \\ + \frac{1}{3}\int\limits_{x_{\pm}}^{1} f_{p}^{d_{V}(n)}(x_1) G_{d}^{{\rm\Lambda}_c}(x_{\pm}/x_1) dx_1,
\end{multline}
\begin{multline}
F_{q_{sea}}^{{\rm\Lambda}_c(n)}(x_{\pm}) = F_{\bar{q}_{sea}}^{{\rm\Lambda}_c(n)}(x_{\pm}) = \\ =
\frac{1}{4}  \int\limits_{x}^{1} f_{p}^{u_{V}(n)}(x_1) [G_{u}^{{\rm\Lambda}_c}(x_{\pm}/x_1)+G_{\bar{u}}^{{\rm\Lambda}_c}(x_{\pm}/x_1)] dx_1 + \\ + \frac{1}{4}\int\limits_{x}^{1} f_{p}^{d_{V}(n)}(x_1) [G_{d}^{{\rm\Lambda}_c}(x_{\pm}/x_1) + G_{\bar{d}}^{{\rm\Lambda}_c}(x_{\pm}/x_1)] dx_1, 
\end{multline}
where
\[
G_{1ud}^{{\rm\Lambda}_c}(x_{\pm}/x_1) = (x_{\pm}/x_1)^{2(\alpha_{R}(0) - \alpha_{N}(0))}
(1-x_{\pm}/x_1)^{\chi},
\]
\[
G_{1uu}^{{\rm\Lambda}_c}(x_{\pm}/x_1) = (x_{\pm}/x_1)^{2}
(1-x_{\pm}/x_1)^{\chi+1},
\]
\[
G_{0ud}^{{\rm\Lambda}_c}(x_{\pm}/x_1) = G_{0uu}^{{\rm\Lambda}_c} = (1-x_{\pm}/x_1)^{\chi+4(1-\alpha_{N}(0))},
\]
\[
G_{u}^{{\rm\Lambda}_c}(x_{\pm}/x_1) = G_{d}^{{\rm\Lambda}_c} = (1-x_{\pm}/x_1)^{\chi+2(\alpha_{R}(0)-\alpha_{N}(0))},
\]
\[
G_{\bar{u}}^{{\rm\Lambda}_c}(x_{\pm}/x_1) = G_{\bar{d}}^{{\rm\Lambda}_c} = (1-x_{\pm}/x_1)^{\chi+2(\alpha_{R}(0)-\alpha_{N}(0))+2(1-\alpha_{R}(0))},
\]
\[
\chi = \lambda-\alpha_{\psi}(0).
\]
\par
Charmed baryons ${\rm\Lambda}_{c}$ have harder spectrum in comparison with charmed mesons in the region $x>0.1$. Fragmentation process of the charmed baryon differs from fragmentation of $D$~mesons since ${\rm\Lambda}_{c}$ baryon consist of three quarks. The diquark fragmentation functions are 
divided into two parts which describe different kinematical regions: $F_{0qq}^{n}(x)$ (central region) and $F_{1qq}^{n}(x)$ (fragmentation region).

The first term in (\ref{equation:phipLc1}) corresponds to direct ${\rm\Lambda}^{+}_{c}$ baryon production and the second one is connected to a pair production of  ${\rm\Lambda}^{+}_{c}/{\rm\Lambda}^{-}_{c}$. 
 In $pp$ interactions ${\rm\Lambda}^{+}_{c}$ baryon can be produced by leading $ud$ diquark, that leads to enhancement of ${\rm\Lambda}^{+}_{c}$ spectra in comparison with ${\rm\Lambda}^{-}_{c}$. Accounting the sea diquark contribution is important only  for antibaryon production in the forward region, while for baryon spectra sea diquarks contribute small practically at all $x$~\cite{Armesto1997}. At $x\rightarrow 1$ dominate direct ${\rm\Lambda}^{+}_{c}$ production in comparison with the pair ${\rm\Lambda}^{+}_{c}/{\rm\Lambda}^{-}_{c}$                production because of diquark function fragmentations in case of pair production is suppressed in forward region by additional term $(1-x_{\pm}/x_1)^{4(1-\alpha_{N}(0))}$.  So here we neglect sea diquarks contribution.

\begin{figure*}
	\centering
     	\includegraphics[trim=0cm 0cm 0cm 0cm, width=0.46\textwidth]{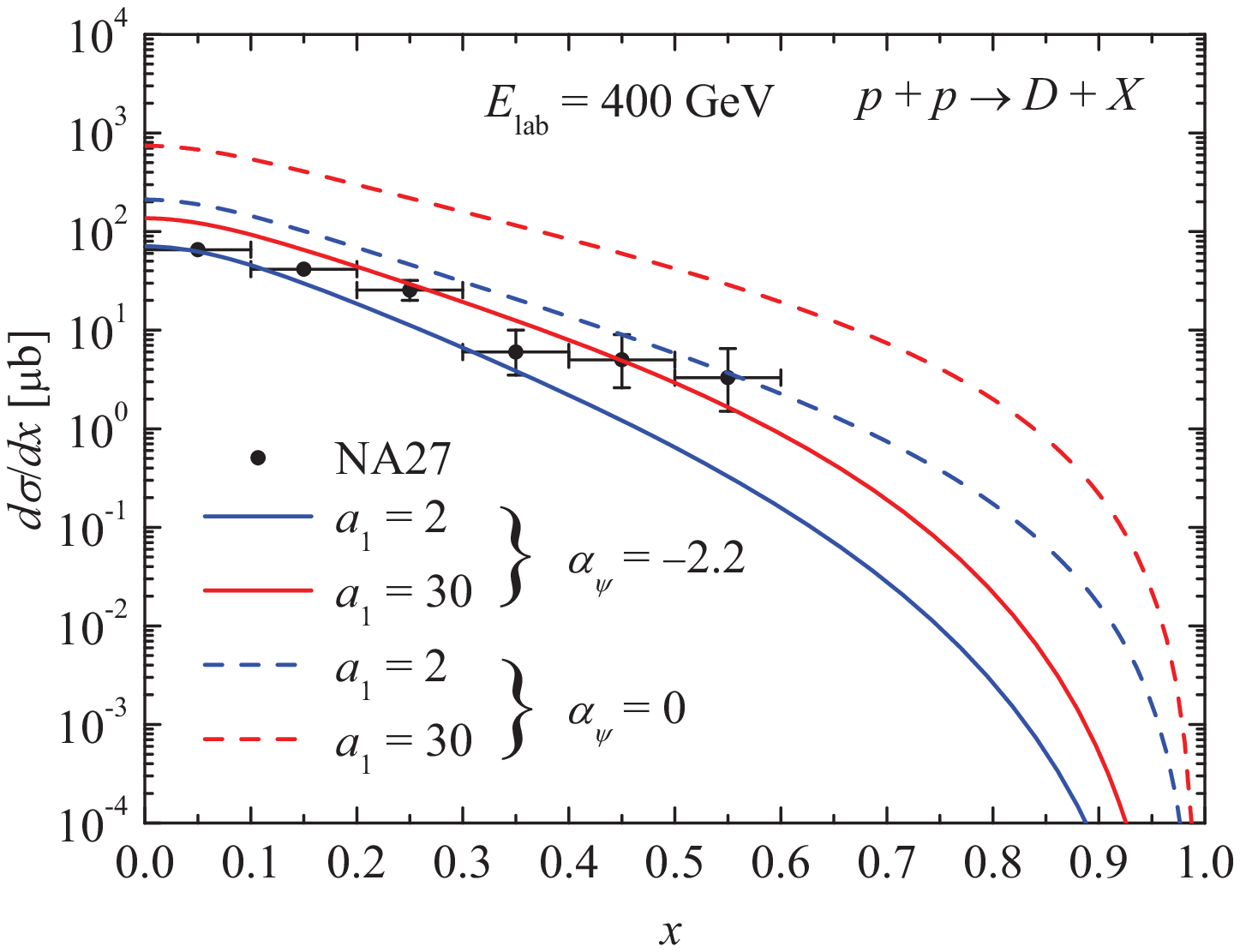}        
	\hspace{3mm}
    	\includegraphics[trim=0cm 0cm 0cm 0cm, width=0.46\textwidth]{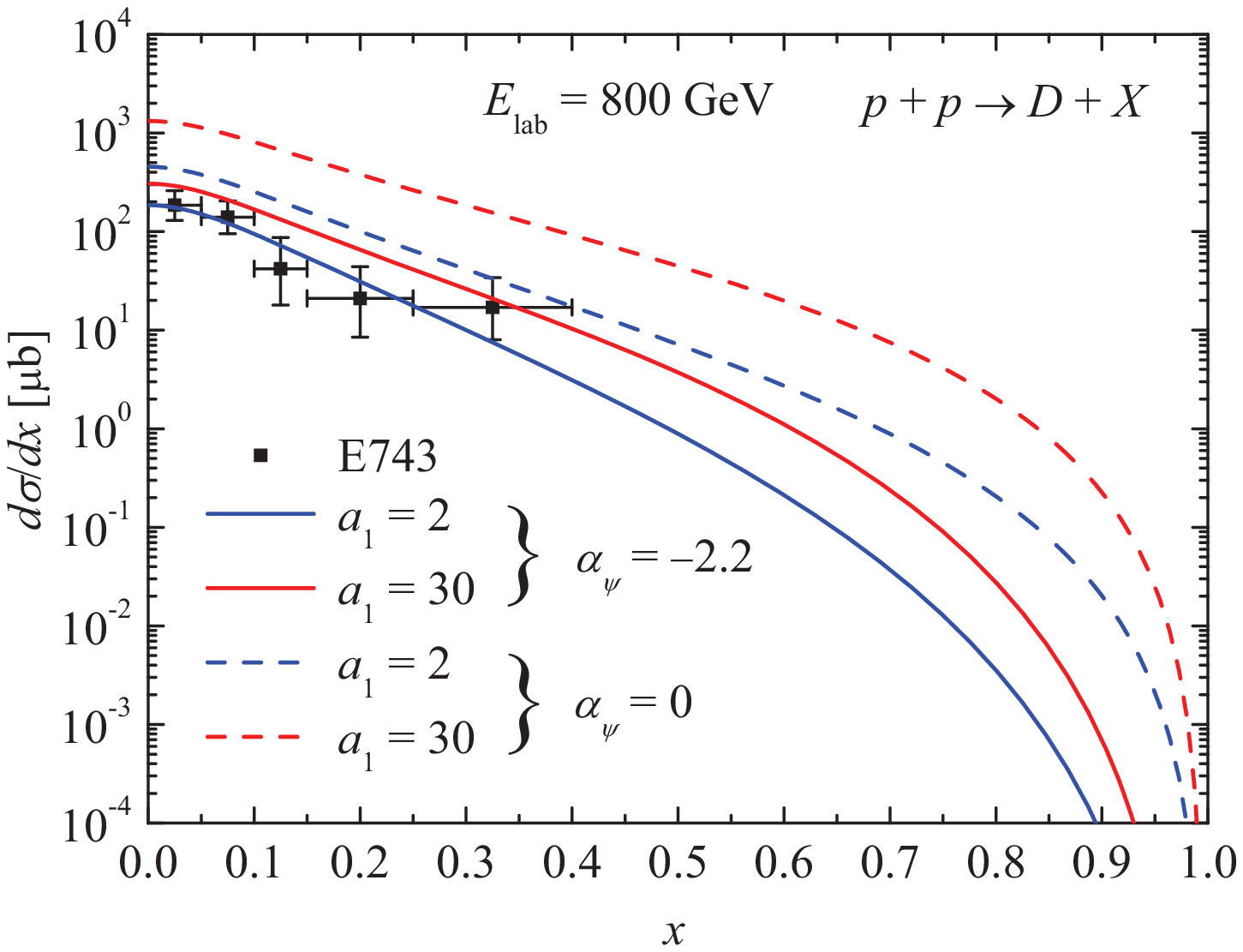} 
	\caption{\label{fig:2}Differential cross sections of~$D/\bar{D}$ mesons production in $pp$ collisions at $E_{\rm lab}=~400$~GeV (left) and $E_{\rm lab}=800$ GeV (right). Experimental data: 
	{$\bullet$}~--~\cite{NA27}; {\tiny $\blacksquare$}~--~\cite{E743}. Same notation for lines as in Fig.~\ref{fig:1}.}
\end{figure*}
\begin{figure*}
	\centering
	\includegraphics[trim=0cm 0cm 0cm 0cm, width=0.46\textwidth]{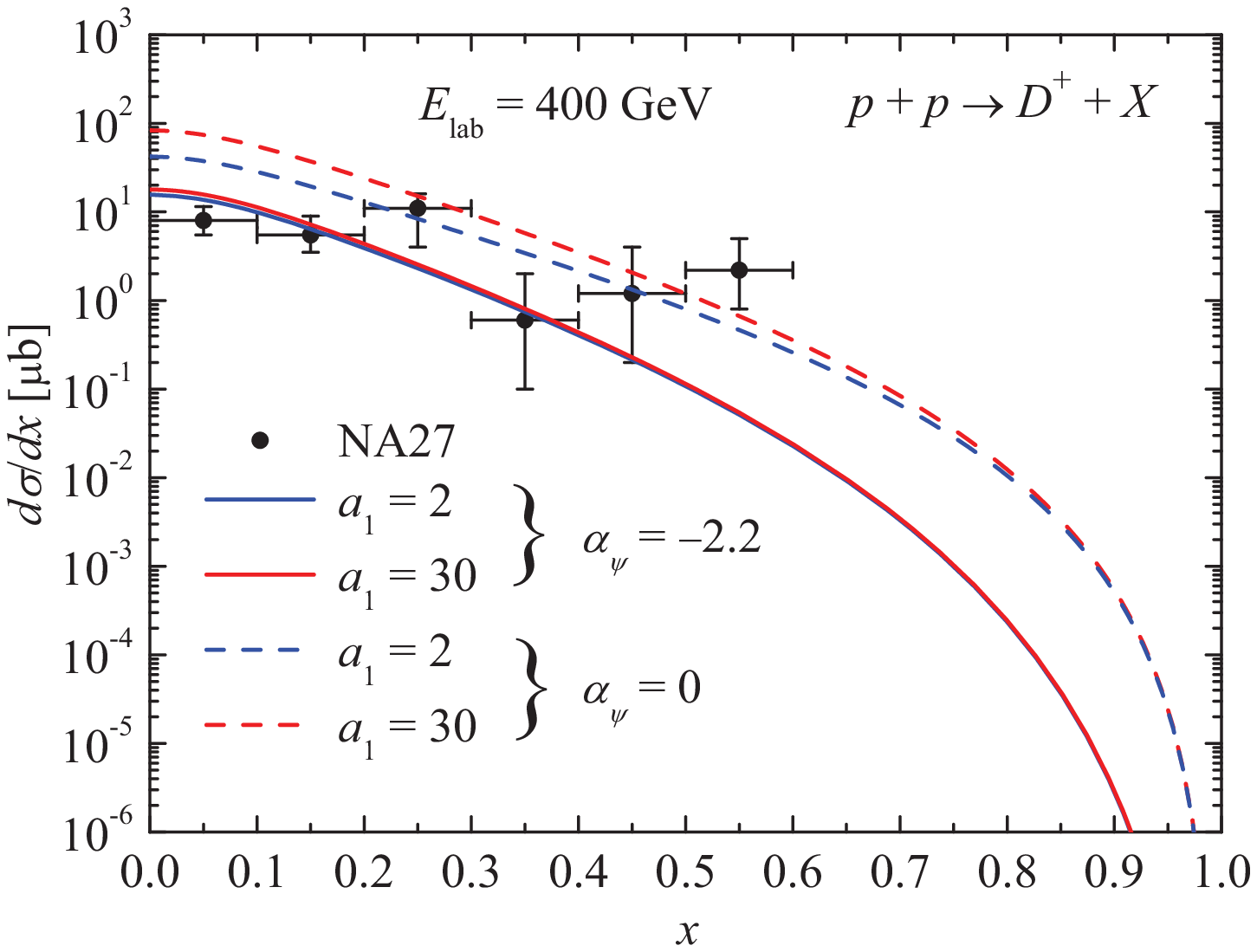}	
	\hspace{3mm}
	\includegraphics[trim=0cm 0cm 0cm 0cm, width=0.46\textwidth]{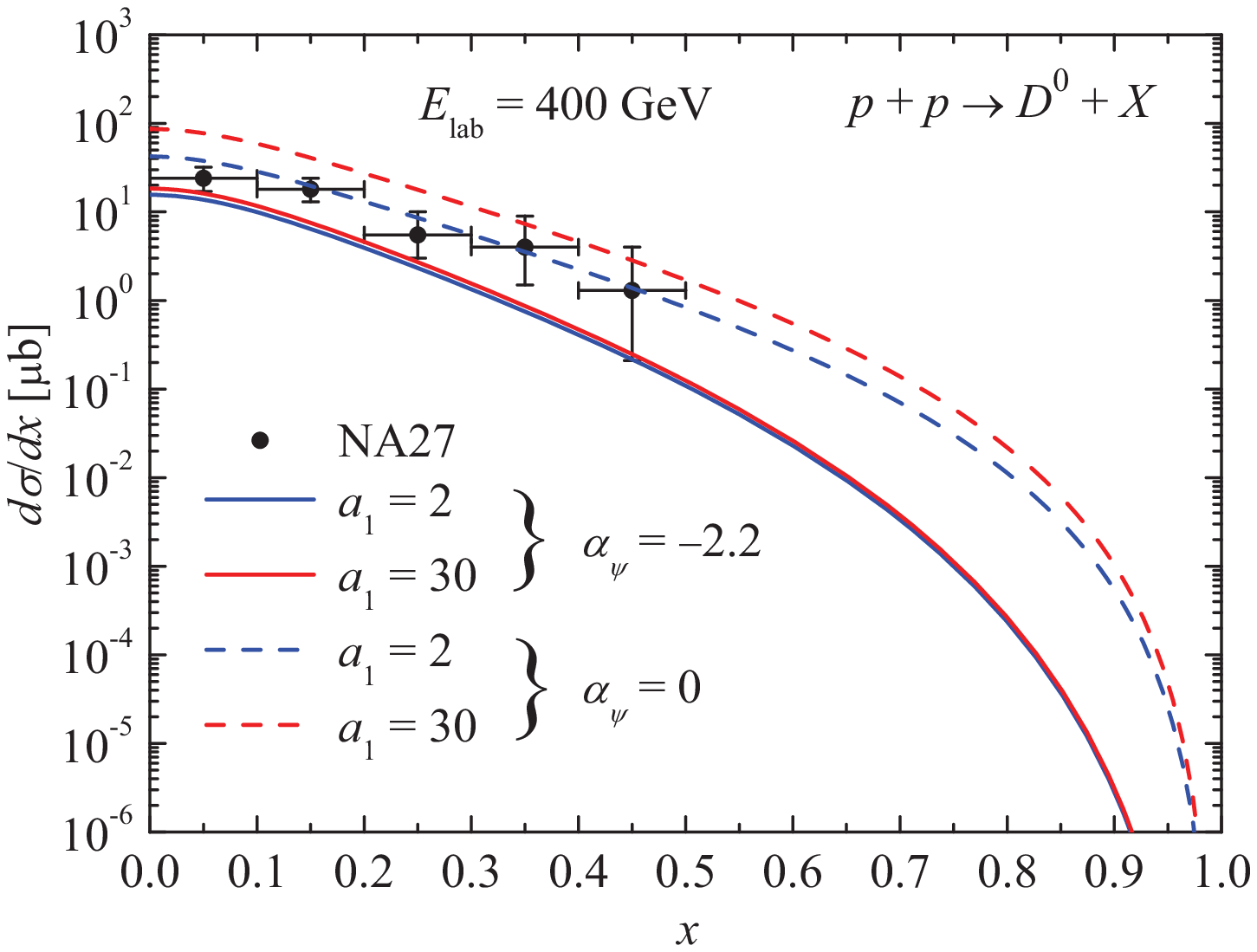}		
		\includegraphics[trim=0cm 0cm 0cm 0cm, width=0.46\textwidth]{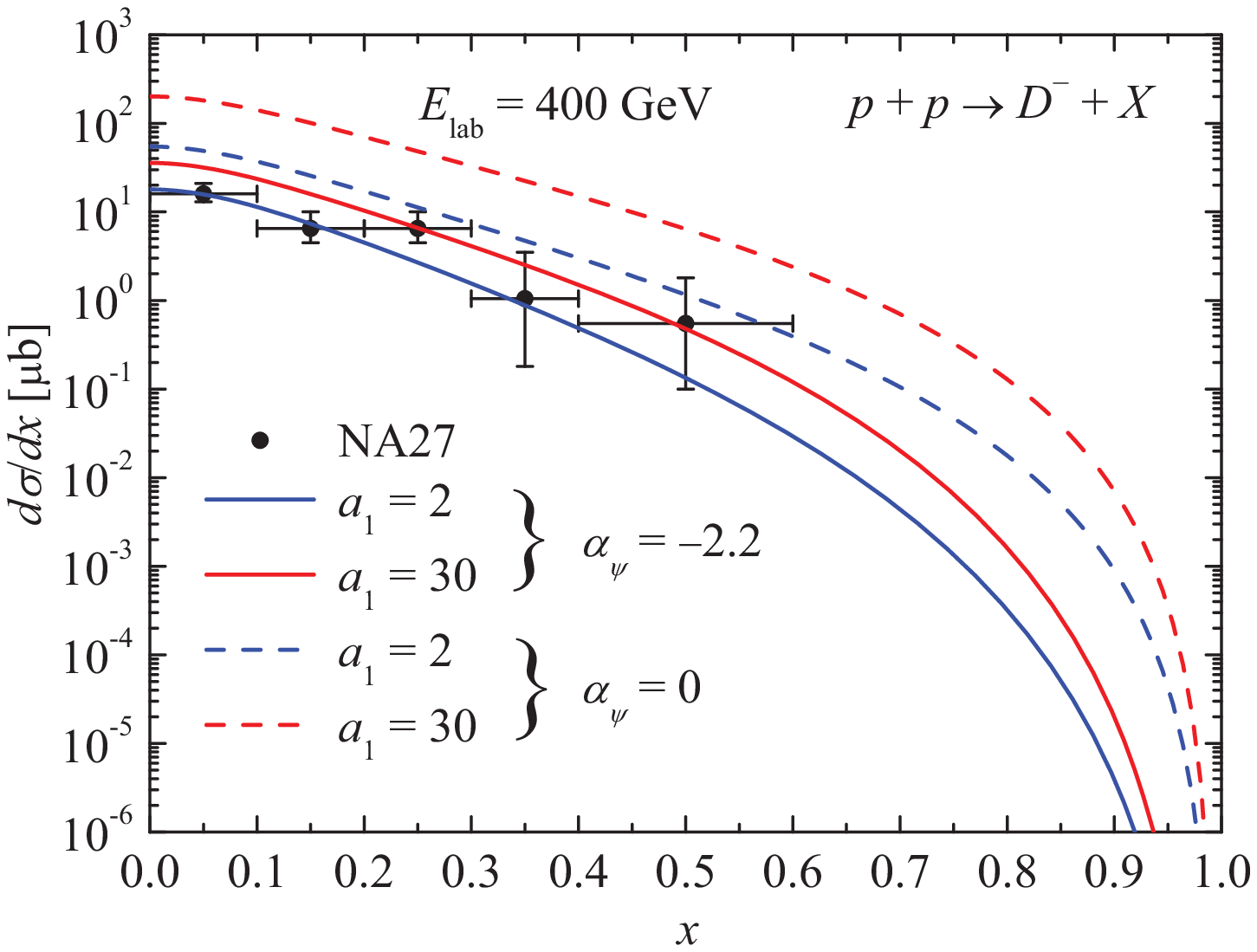}
	\hspace{3mm}
			\includegraphics[trim=0cm 0cm 0cm 0cm, width=0.46\textwidth]{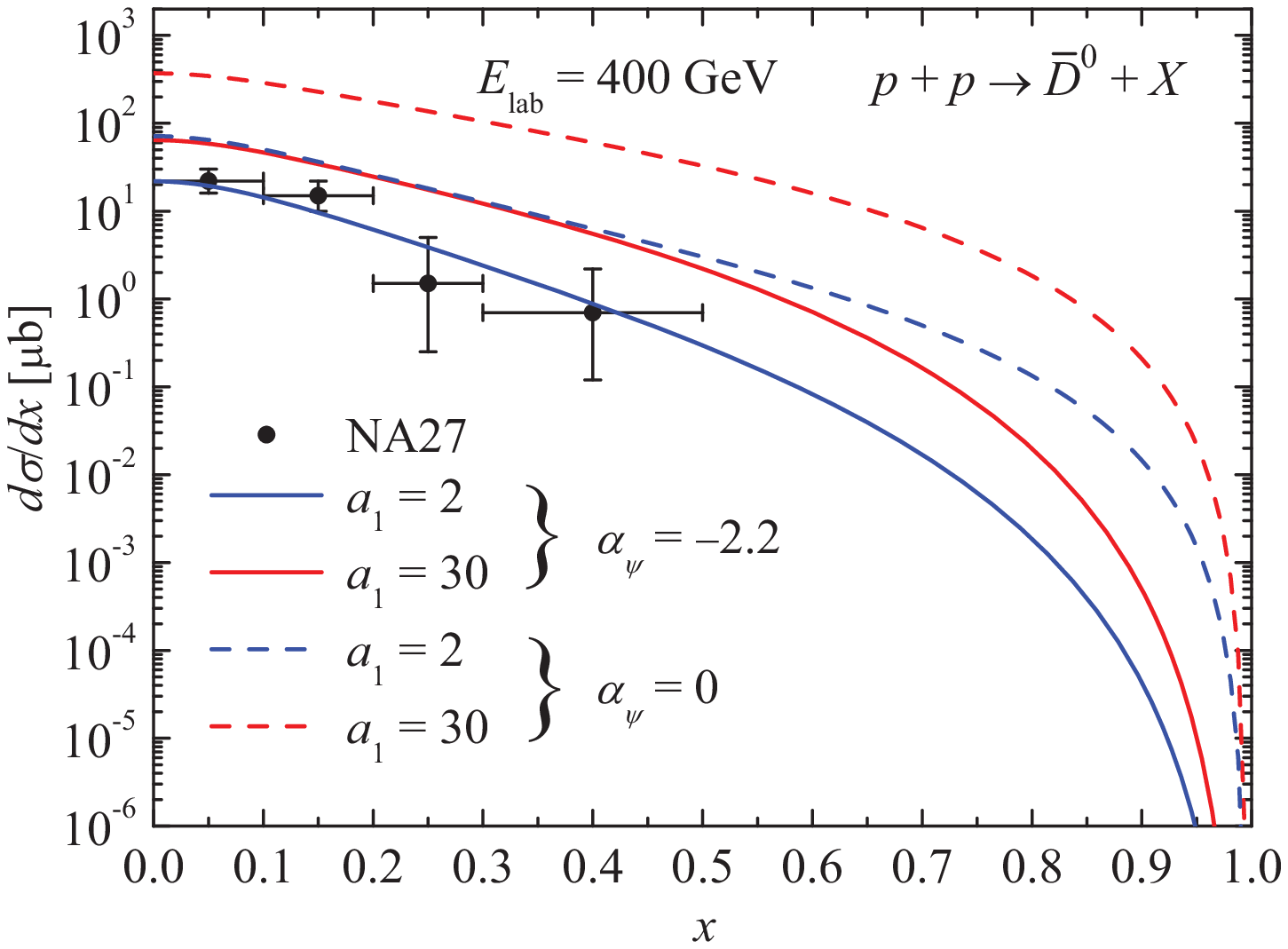}
	\caption{\label{fig:3} Differential cross sections of each type of $D$~mesons ($D^{+}, D^{-}, D^{-}, \bar{D}^0$) production in $pp$ collisions at $E_{\rm lab}=400$~GeV. Lines plot the result of QGSM  calculations with~$\alpha_{\psi}(0)=-2.2$ (solid lines) and $\alpha_{\psi}(0)=0$ (dash). Experimental data are from~\cite{NA27}.}
\end{figure*}
\par  


The distribution functions of charmed particles in (\ref{equation:phip}) contain free parameters that cannot be calculated within the framework of the quark-gluon string model, and their values should be found from a comparison with experiments. The intercept~$\alpha_{\psi}(0)$ of the poorly studied $c\bar{c}$~-trajectory noticeably affects the cross sections of $D$~meson production.
Two values of $\alpha_{\psi}(0)$ have been used by QGSM authors \cite{KaidalovPiskunova_charm,Kaidalov03_YaF}: $\alpha_{\psi}(0)=-2.2$, obtained from the mass spectrum on the assumption of linear Regge trajectory,  and  $\alpha_{\psi}(0)=0$ (nonlinear trajectory), derived  from the perturbative calculations. Basing on experimental data on ${\rm\Lambda}_{c}$  production, they consider \cite{Kaidalov03_YaF} nonperturbative value $\alpha_{\psi}(0)=-2.2$ as preferable one. 
If the Regge trajectory~$\alpha_{\psi}(t)$ is linear (similar to light hadrons), then the $x$-distributions of charmed particles become softer in comparison with the case of $\alpha_{\psi}(0)=0$. 
\par 
The coefficient $a_{1}$ provides an unified description of the kinematic regions $x\rightarrow0$ and $x\rightarrow1$ in the case of leading fragmentation (when the valence quarks take part in the fragmentation). Now there are no clear arguments for choice of its value, and different authors 
apply various values among which two extreme values may be chosen: $a_{1}=2$~\cite{Arakelyan} and $a_{1}=30$~\cite{KaidalovPiskunova_charm}. New measurements of the total cross sections of charmed meson production at high energies in the experiments ALICE~\cite{ALICE_Abelev,ALICE_Adam,ALICE_Acharya} and ATLAS~\cite{ATLAS} allow a check of the QGSM predictions for extreme values of the parameter $a_{1}$.
\par
The parameter $a^{h}$ in (\ref{equation:phip}) concerns the charmed particles number density in the central region of the inclusive spectra and in an obvious way affects on the cross sections in (\ref{equation:inclusiveCS}). For the $D$ and ${\rm\Lambda}^{+}_{c}$ particles, we use the values  from Refs.~\cite{KaidalovPiskunova_charm,Arakelyan}: 
$a^{D}=7.0\cdot10^{-4}$, 
$a_0^{{\bar{\rm\Lambda}}_{c}}=7\cdot10^{-4}$,  
$a^{{\rm\Lambda}_{c}}_{1}=0.12$ (for~$\alpha_{\psi}(0)=-2.2$) and $a^{{\rm\Lambda}_{c}}_{1}=0.02$ (for~$\alpha_{\psi}(0)=0$).  The value $a^{D}=1.0\cdot10^{-3}$  is also acceptable for LHCb experimental data.

The results of calculation of the cross sections of $D$~meson production in 
$pp$~collisions in comparison with experimental data are shown in Figs.~\ref{fig:1}--\ref{fig:3}. The total cross section of $D/\bar{D}$~mesons production in $pp$~collisions as a function of center-of-mass energy is calculated in the QGSM for four sets of free parameters (Fig.~\ref{fig:1}). 
\begin{figure} [t]
	\centering
		\includegraphics[trim=0cm 0cm 0cm 0cm, width=0.46\textwidth]{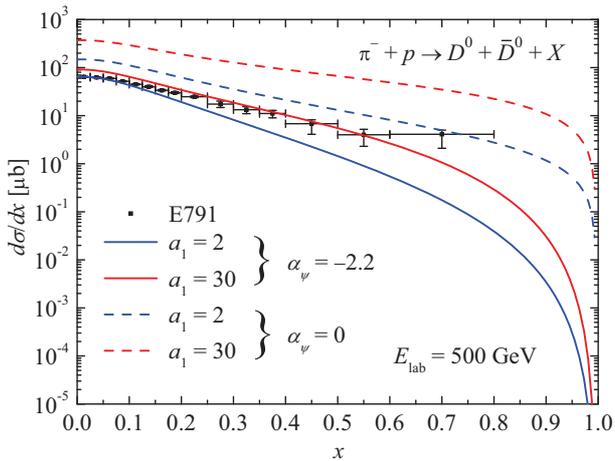}
	\caption{\label{fig:7}Differential cross sections of $D^{0}/\bar{D}^{0}$~mesons production in $\pi p$ collisions at $E_{\rm lab}=500$~GeV. Experimental data are from~\cite{E791}. Same notation for lines as in Fig.~\ref{fig:3}.}
\end{figure}
\begin{figure} [t] 
		\centering
		\includegraphics[trim=0cm 0cm 0cm 0cm, width=0.46\textwidth]{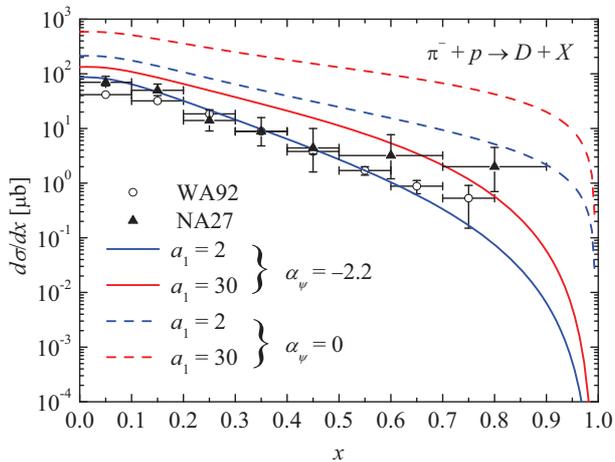}
	\caption{\label{fig:4} Differential cross sections of~$D/\bar{D}$ mesons production in $\pi p$ collisions calculated in the QGSM at $E_{\rm lab}=360$~GeV. Same notation for lines as in Fig.~\ref{fig:3}. 
Experimental data for $E_{\rm lab}=350$~GeV (WA92)~\cite{WA92}, and for $E_{\rm lab}=360$~GeV (NA27)~\cite{NA27_pip}.}
\end{figure}
Here the experimental results in a wide energy  
range~\cite{NA32,E769,NA16,NA27,E743,E653,HERA-B,PHENIX,STAR} including LHC 
measurements~\cite{ALICE_Abelev,ALICE_Adam,ALICE_Acharya,ATLAS} are  presented.
Calculation with~$\alpha_{\psi}(0)=0$ and~$a_{1}=30$ does not agree with 
experimental data at $\sqrt{s}<1$~TeV, while the calculations with~$\alpha_{\psi}(0)=-2.2$  are in close agreement with the measurements in a wide energy range. At low energies the cross sections calculated with~$\alpha_{\psi}(0)=-2.2$ for extreme values $a_{1}=2$ and $a_{1}=30$ differ by a factor~$2-5$, but the influence of  the parameter~$a_{1}$ tends to diminish  with energy  and becomes negligible at high energies ($\sqrt{s}>1$~TeV).
\par
Figure~\ref{fig:2} represents the calculated differential cross sections of $D$ mesons production at the laboratory energies 400 GeV and 800 GeV in comparison with the measurements of experiments NA27~\cite{NA27} and E743~\cite{E743}.

\begin{figure*} 
	\centering
			\includegraphics[trim=0cm 0cm 0cm 0cm, width=0.46\textwidth]{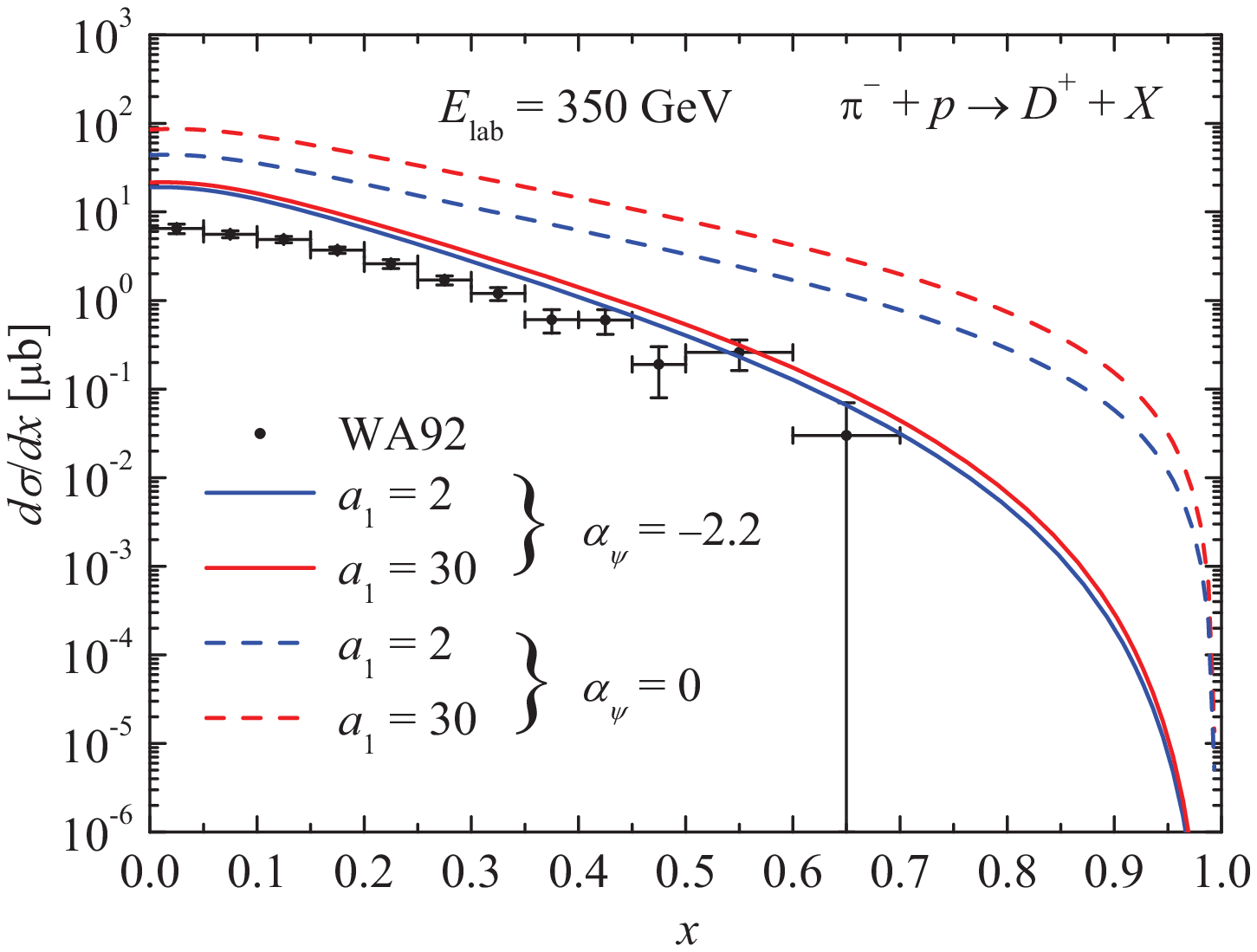}
	\hspace{3mm}
	\includegraphics[trim=0cm 0cm 0cm 0cm, width=0.46\textwidth]{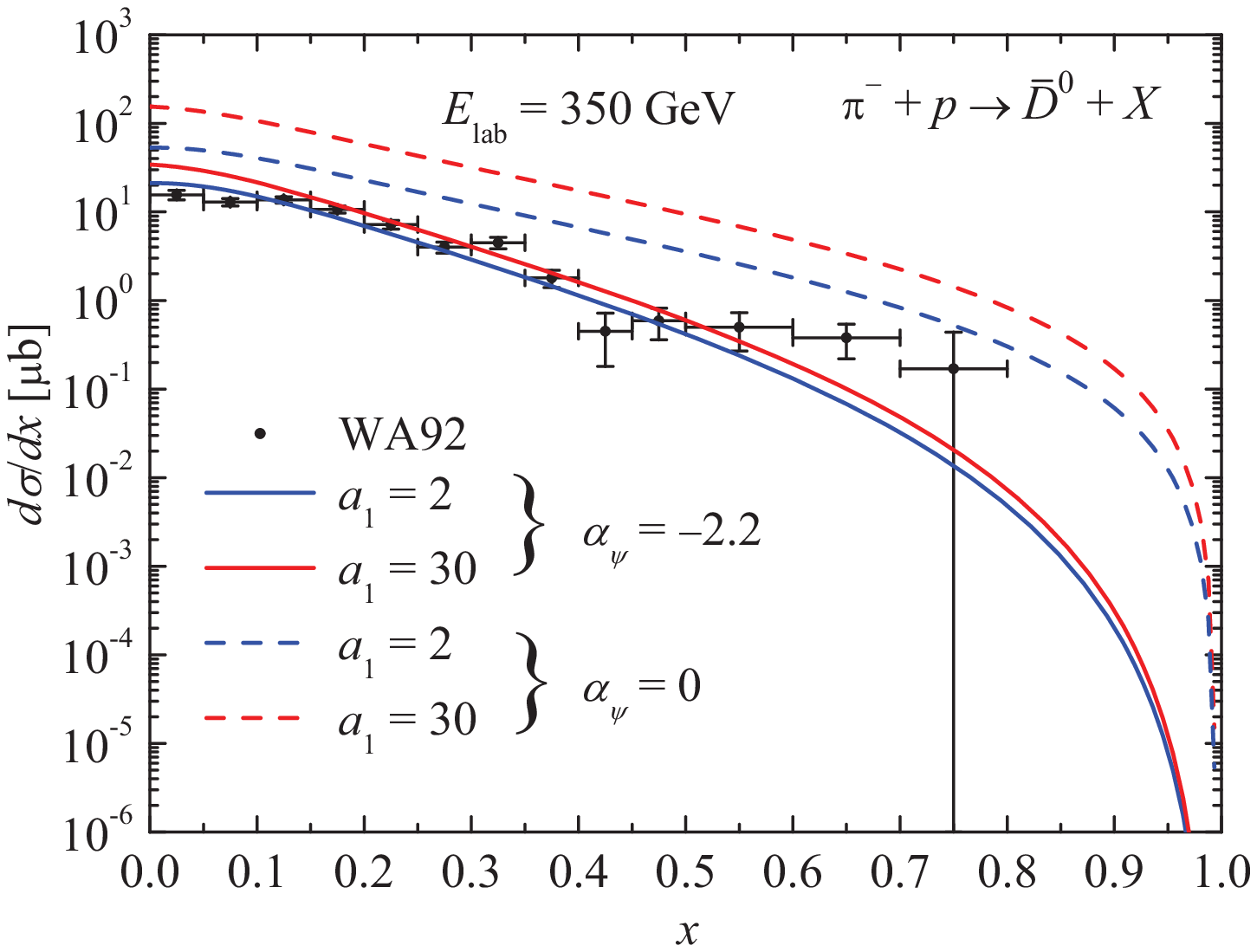}
		\includegraphics[trim=0cm 0cm 0cm 0cm, width=0.46\textwidth]{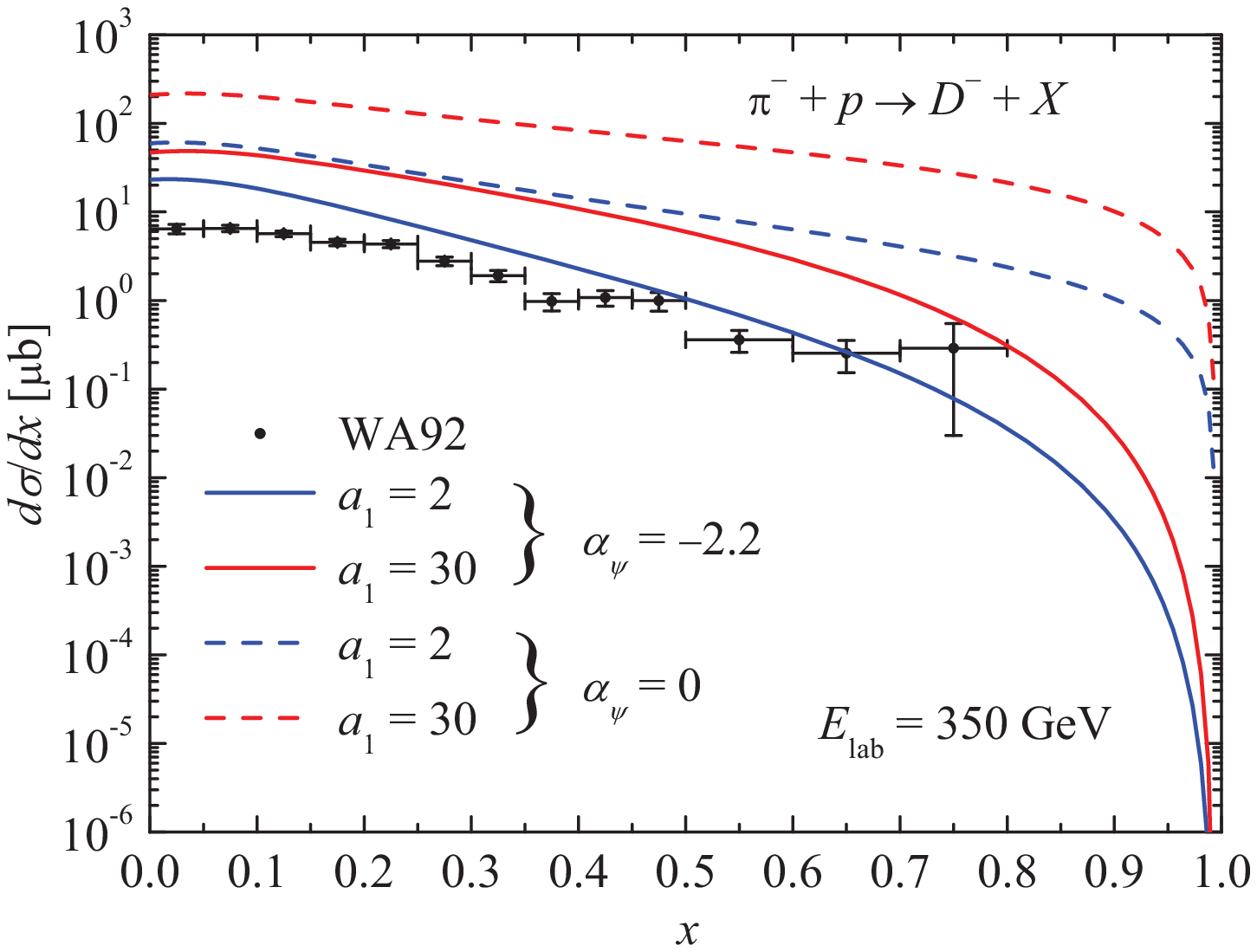}
	\hspace{3mm}
 \includegraphics[trim=0cm 0cm 0cm 0cm, width=0.46\textwidth]{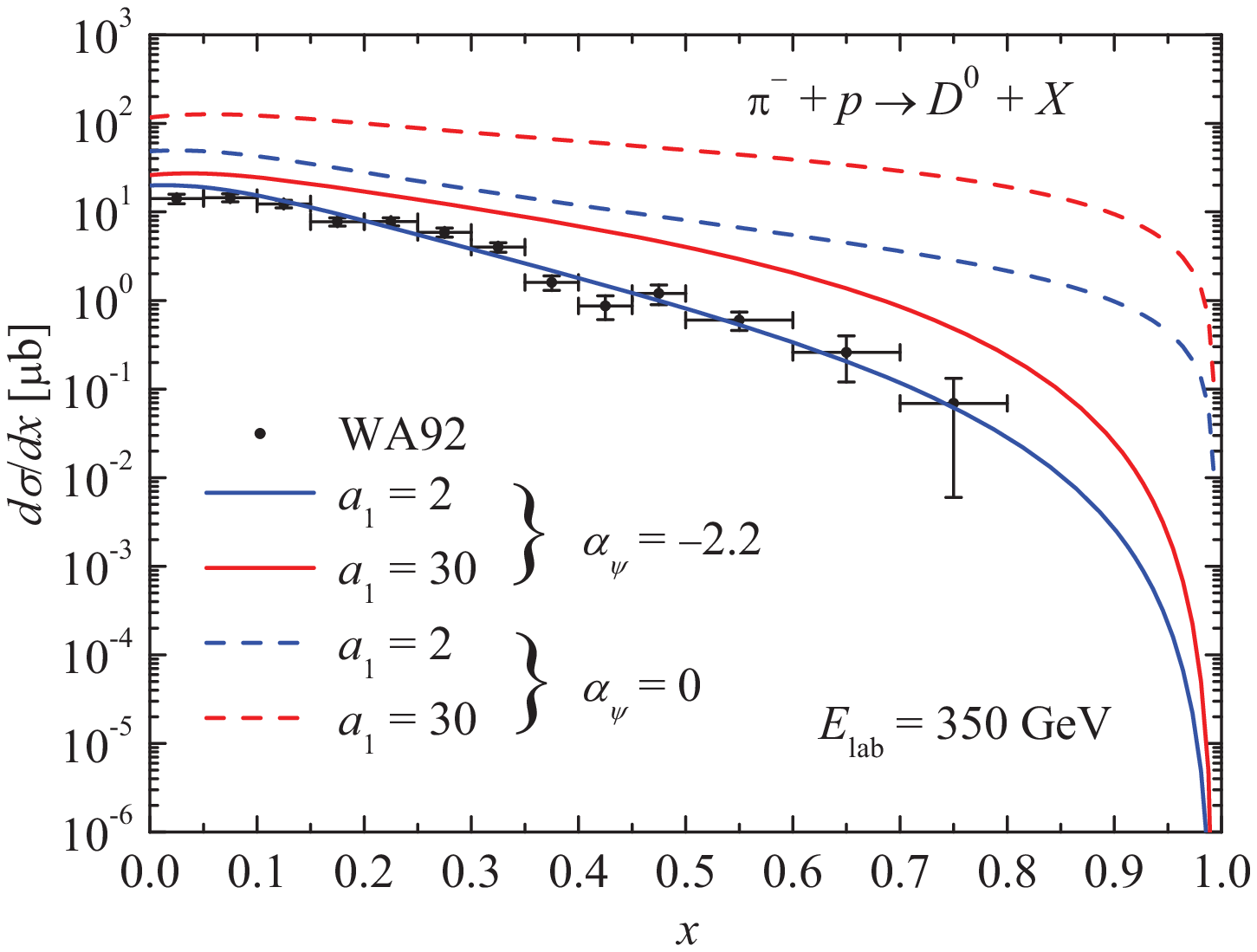}		
	\caption{\label{fig:5} Differential cross sections of each type of $D$~mesons ($D^{+}, D^{-}, D^{-}, \bar{D}^0$) production in $\pi p$ collisions at $E_{\rm lab}=350$~GeV. QGSM calculations for~$\alpha_{\psi}(0)=-2.2$ (solid lines) and $\alpha_{\psi}(0)=0$ (dash). Points are experimental data  from~\cite{WA92}.}
\end{figure*}
\begin{figure*} 
	\centering
		\includegraphics[trim=0cm 0cm 0cm 0cm, width=0.46\textwidth]{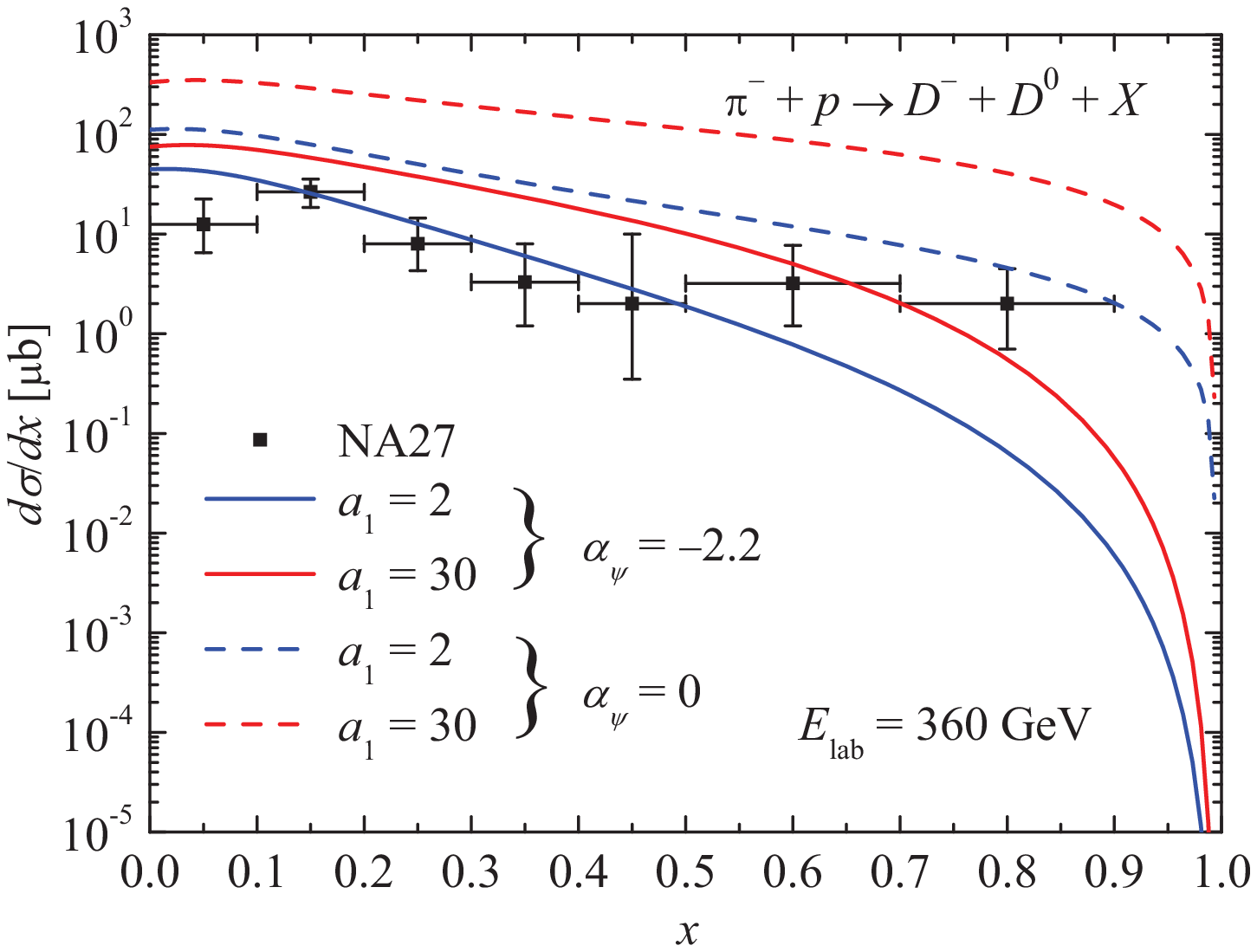}
	\hspace{3mm}
		\includegraphics[trim=0cm 0cm 0cm 0cm, width=0.46\textwidth]{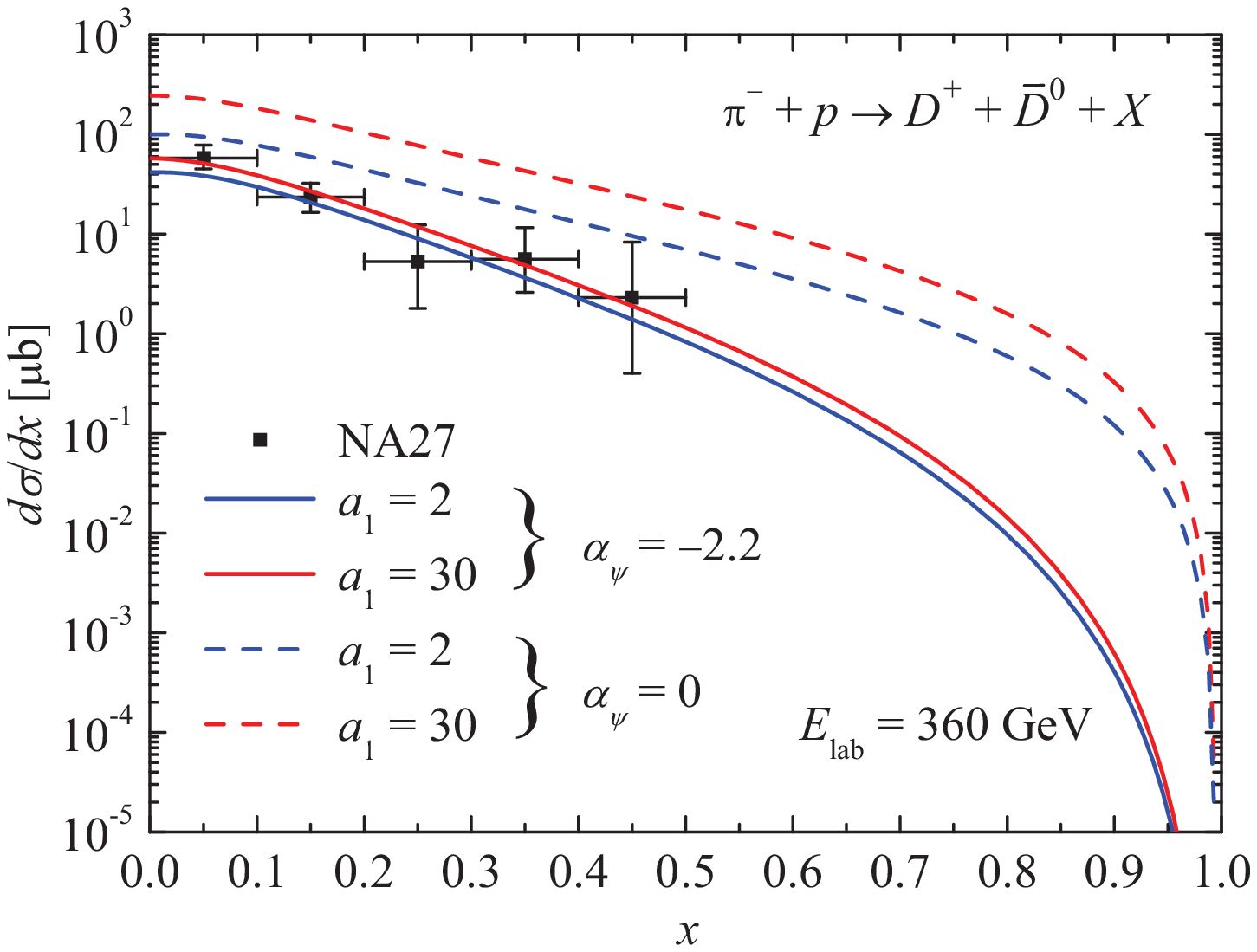}
	\caption{\label{fig:6}Differential cross sections of "leading" $D^-/D^0$ mesons (left) and "nonleading" $D^+/\bar{D}^0 $ mesons (right) in $\pi p$ collisions at $E_{\rm lab}=~360$~GeV. Calculations for~ $\alpha_{\psi}(0)=-2.2$ (solid line) and $\alpha_{\psi}(0)=0$ (dash). Points are the data of experiment~\cite{NA27_pip}.}
\end{figure*}  
\begin{figure*} 
	\centering
	\includegraphics[trim=0cm 0cm 0cm 0cm, width=0.46\textwidth]{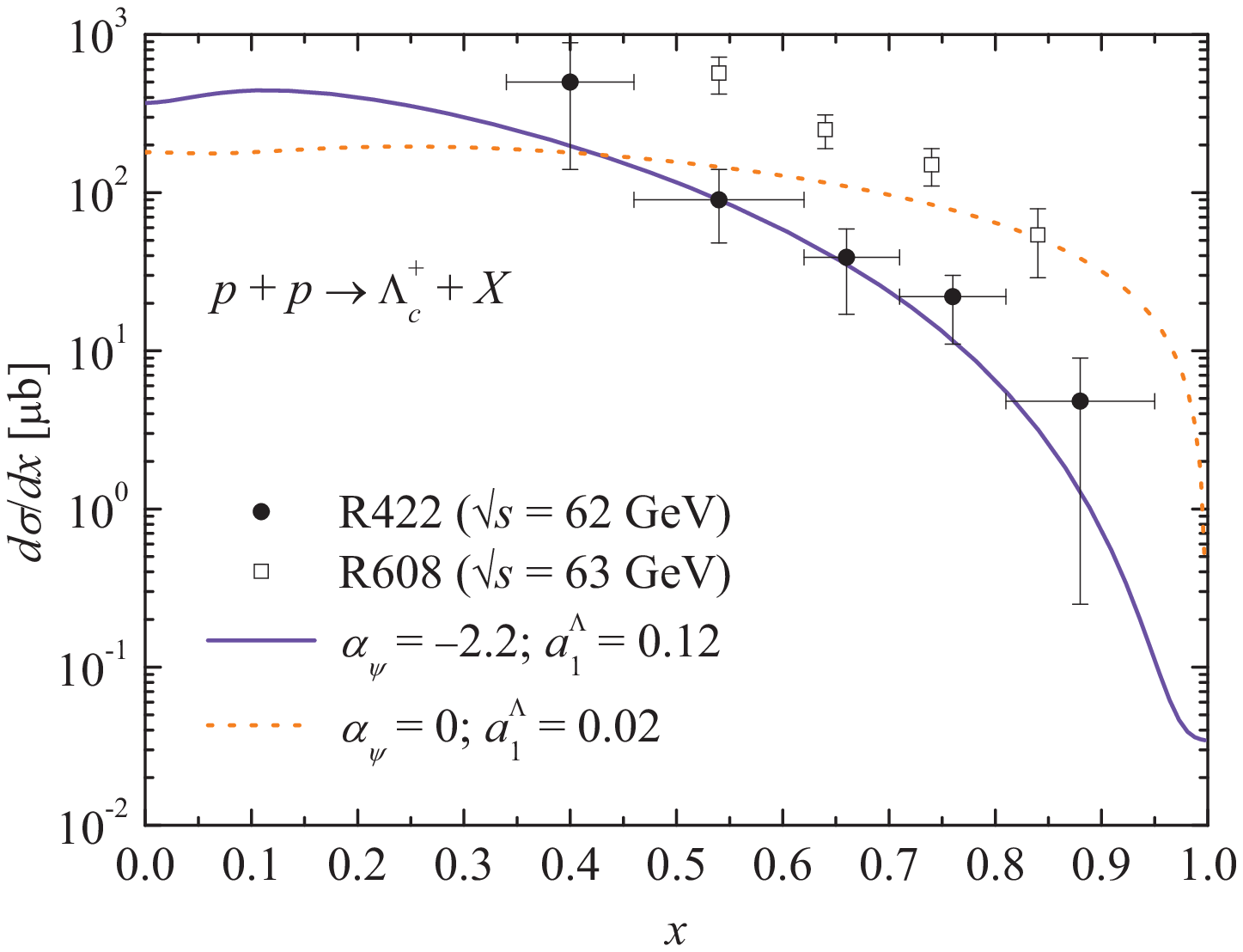}
	\hspace{3mm}
	\includegraphics[trim=0cm 0cm 0cm 0cm, width=0.46\textwidth]{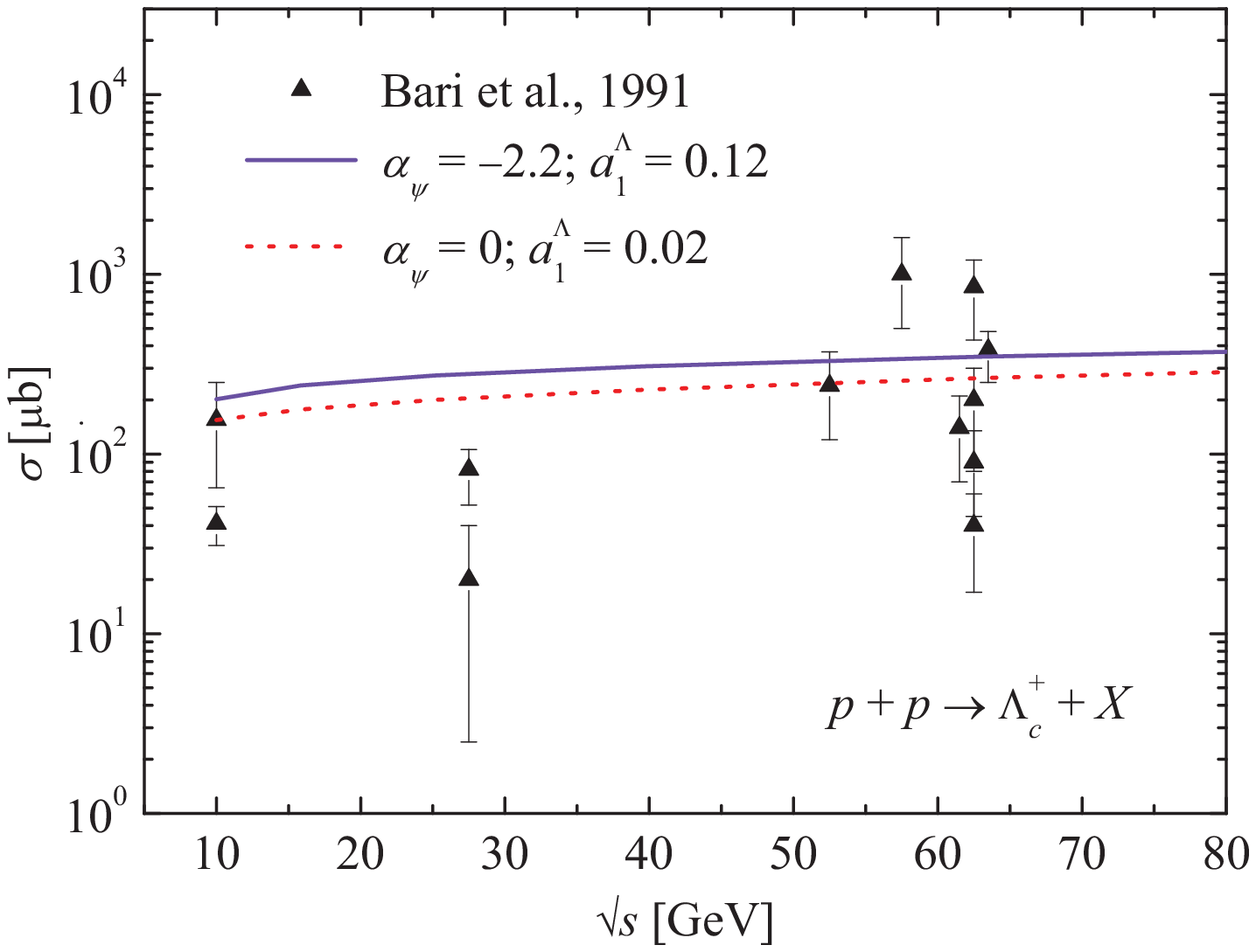}
	\caption{\label{fig:8}Differential (left) and total (right) cross sections of ${\rm\Lambda}^{+}_{c}$ baryon production in $pp$ collisions: calculations for~$\alpha_{\psi}(0)=-2.2$, $a^{{\rm\Lambda}_{c}}_{1}=0.12$~(solid line) and $\alpha_{\psi}(0)=0$, $a^{{\rm\Lambda}_{c}}_{1}=0.02$~(dotted line). Experimental data:  $\bullet$, $\blacktriangle$~--~\cite{Lambda_Bari}; {\tiny $\square$}~--~\cite{Lambda_Chauvat}.}
\end{figure*}
\begin{figure*} 
	\centering
	\includegraphics[trim=0cm 0cm 0cm 0cm, width=0.46\textwidth]{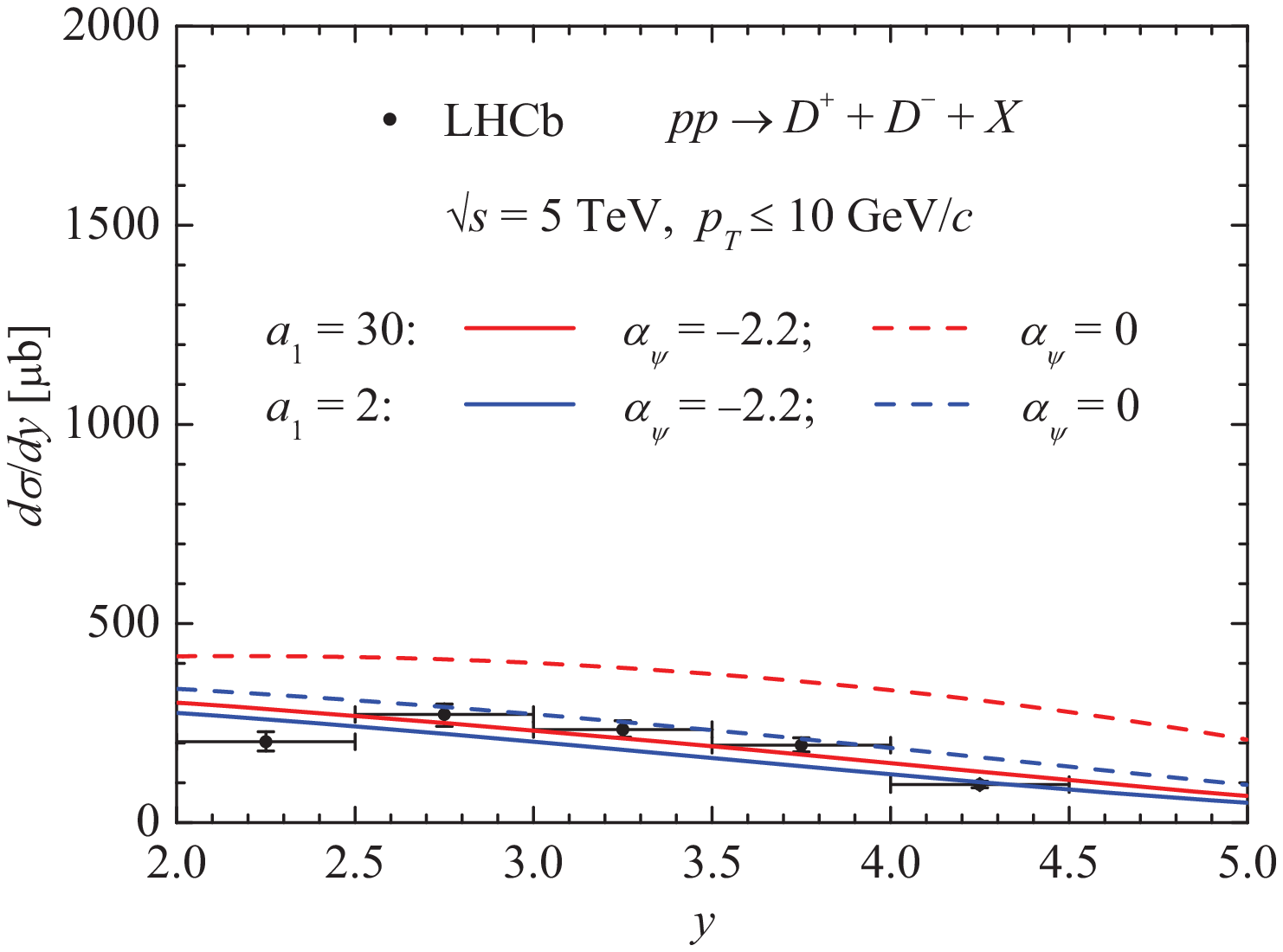}
	\hspace{3mm}
	\includegraphics[trim=0cm 0cm 0cm 0cm, width=0.46\textwidth]{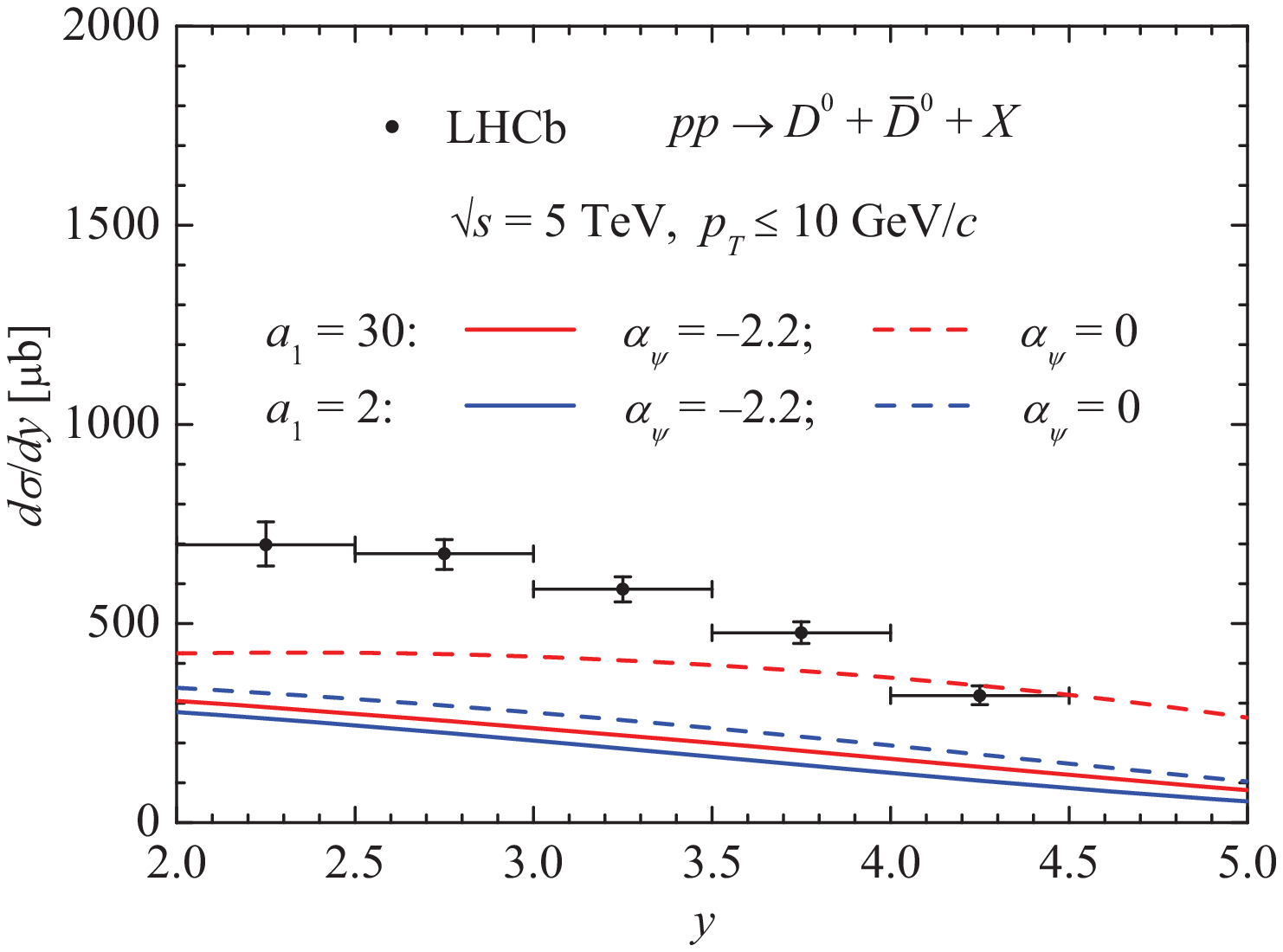}
	\caption{\label{fig:9}Differential cross sections of charged $D$ mesons (left) and neutral $D$ mesons (right) in $pp$ collisions at $\sqrt{s}=5$ TeV. Data are from LHCb experiment~\cite{LHCb5TeV}. QGSM calculations: $\alpha_{\psi}(0)=-2.2$ (solid lines), $\alpha_{\psi}(0)=0$ (dash lines);  $a_{1}=2$ (blue bottom), $a_{1}=30$  (red).} 
\end{figure*}
\begin{figure*} 
	\centering
	\includegraphics[trim=0cm 0cm 0cm 0cm, width=0.46\textwidth]{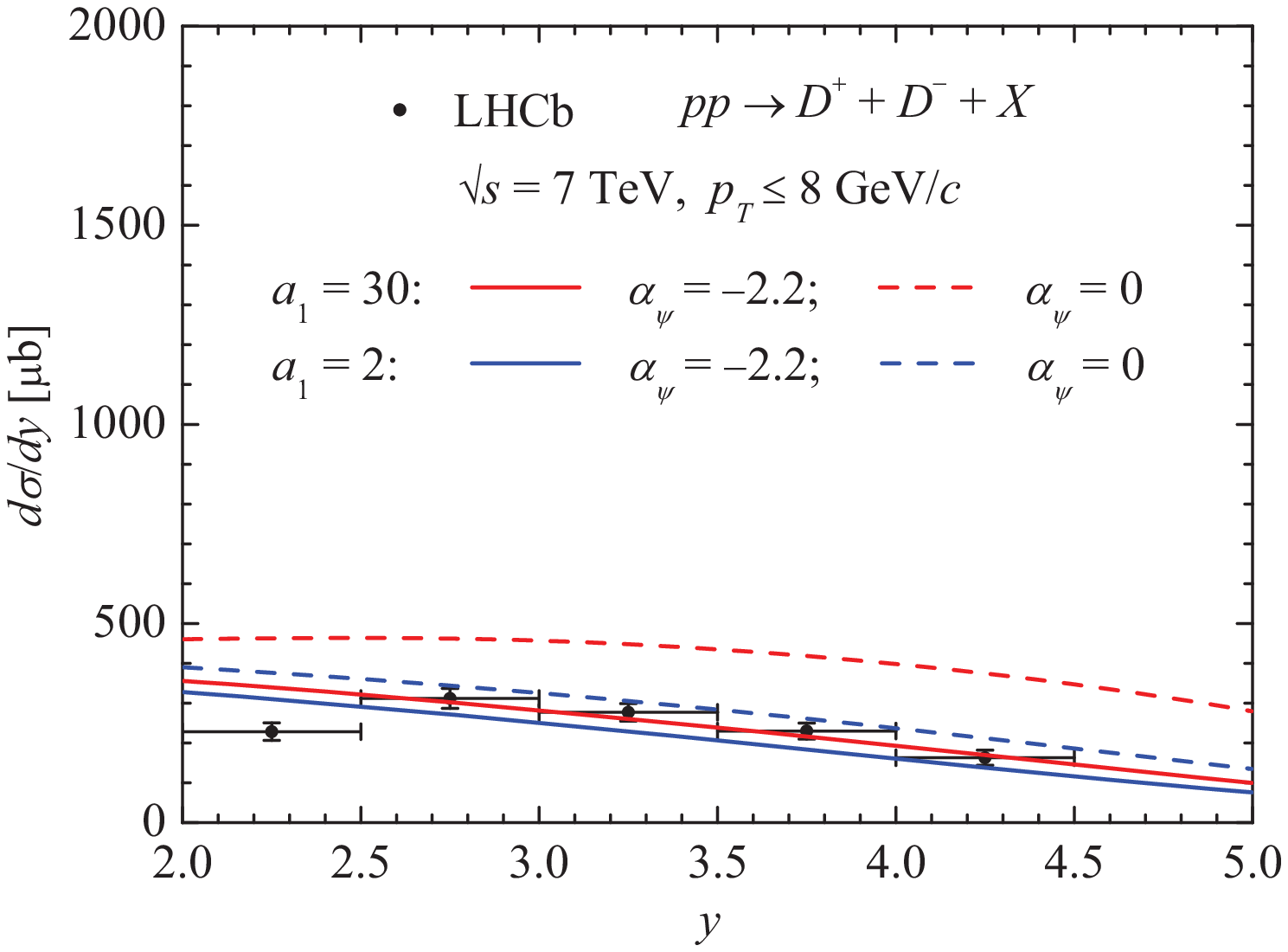}
	\hspace{3mm}
		\includegraphics[trim=0cm 0cm 0cm 0cm, width=0.46\textwidth]{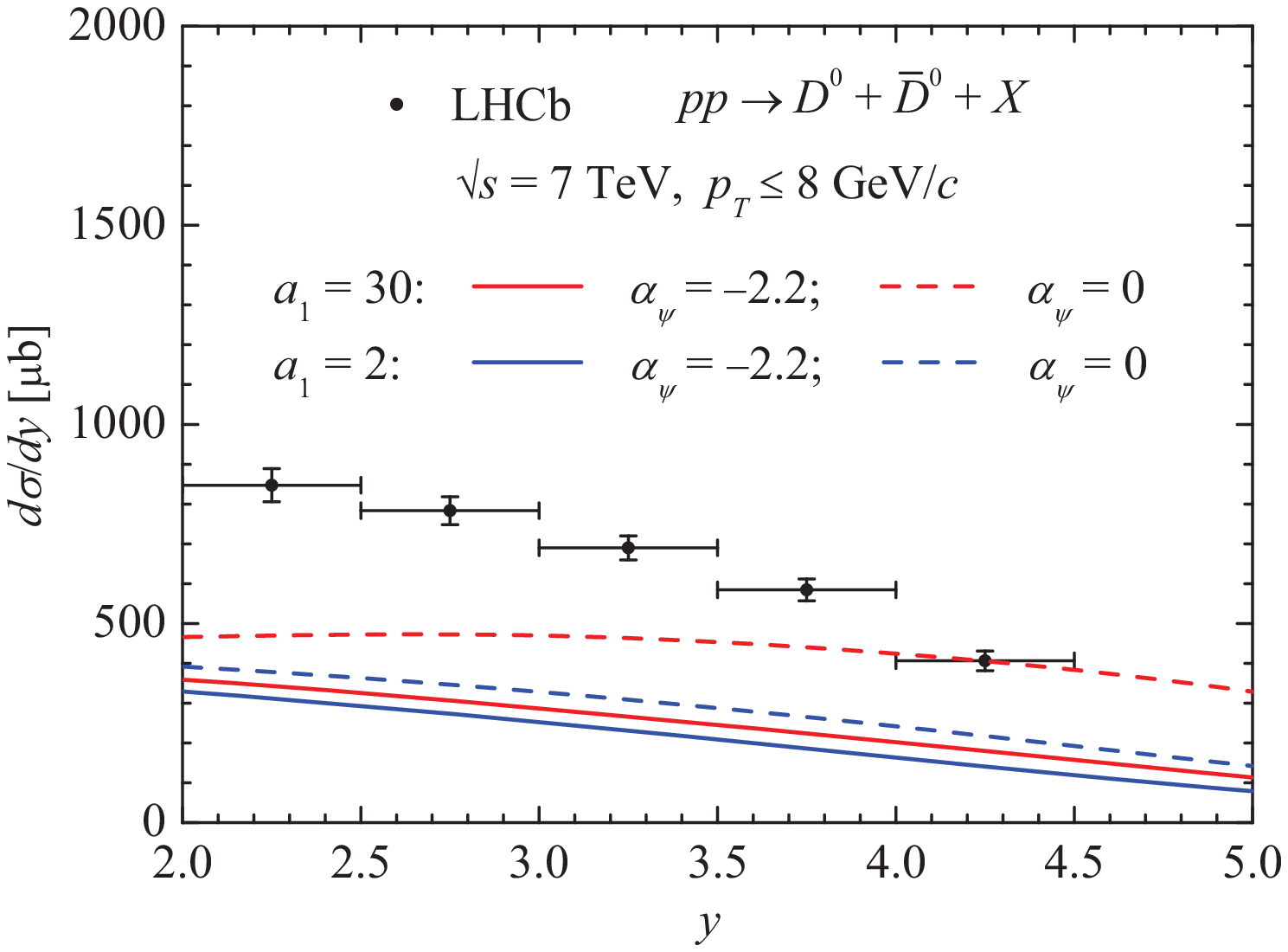}	
	\caption{\label{fig:10}Differential cross sections of charged $D$ mesons (left) and neutral $D$ mesons (right) in $pp$ collisions at $\sqrt{s}=7$ TeV. Data are from LHCb experiment~\cite{LHCb7TeV}. The same notation for lines as in Fig.~\ref{fig:9}.}
\end{figure*}
\begin{figure*} 
	\centering
			\includegraphics[trim=0cm 0cm 0cm 0cm, width=0.46\textwidth]{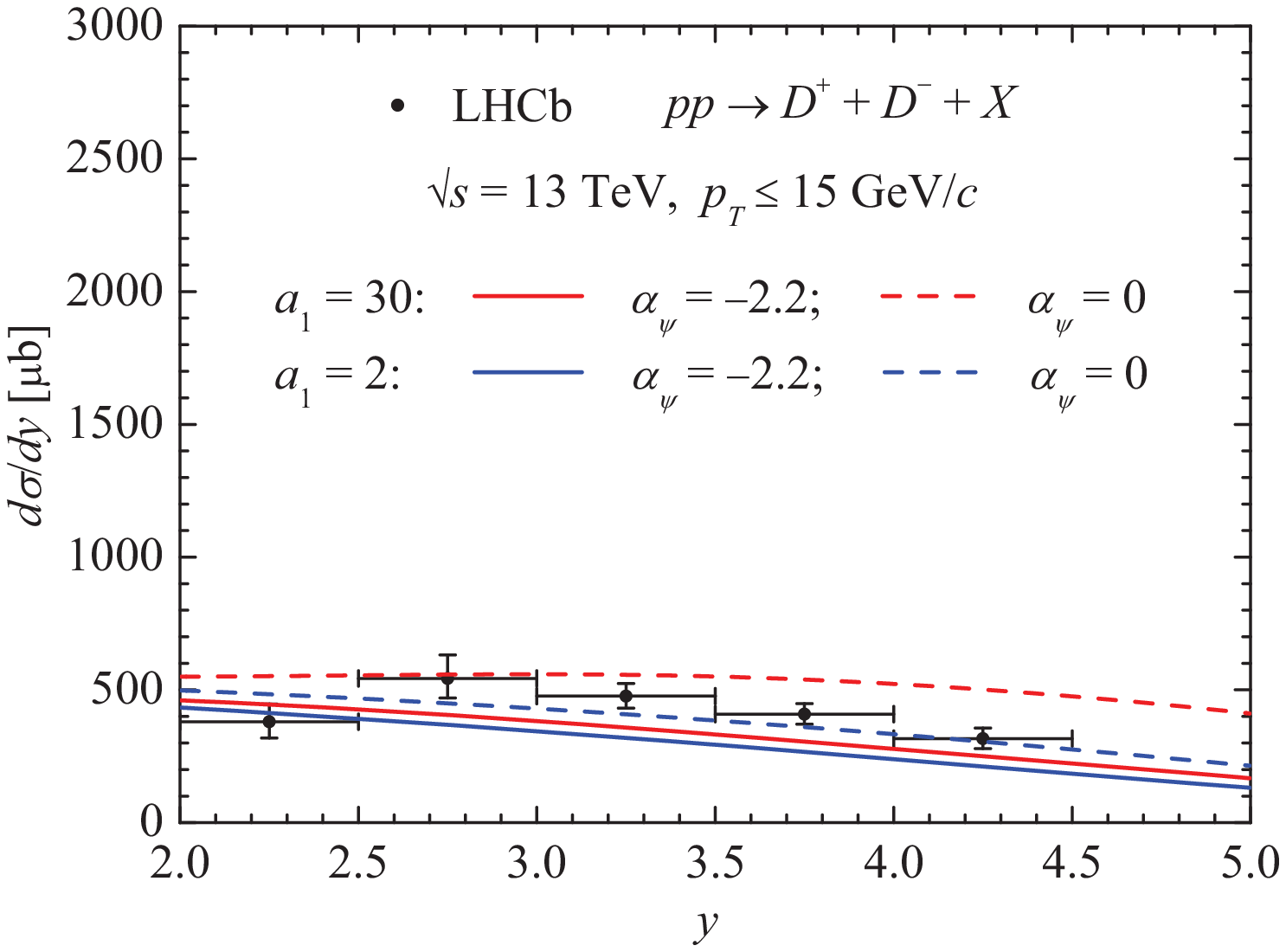}
	\hspace{3mm}
		\includegraphics[trim=0cm 0cm 0cm 0cm, width=0.46\textwidth]{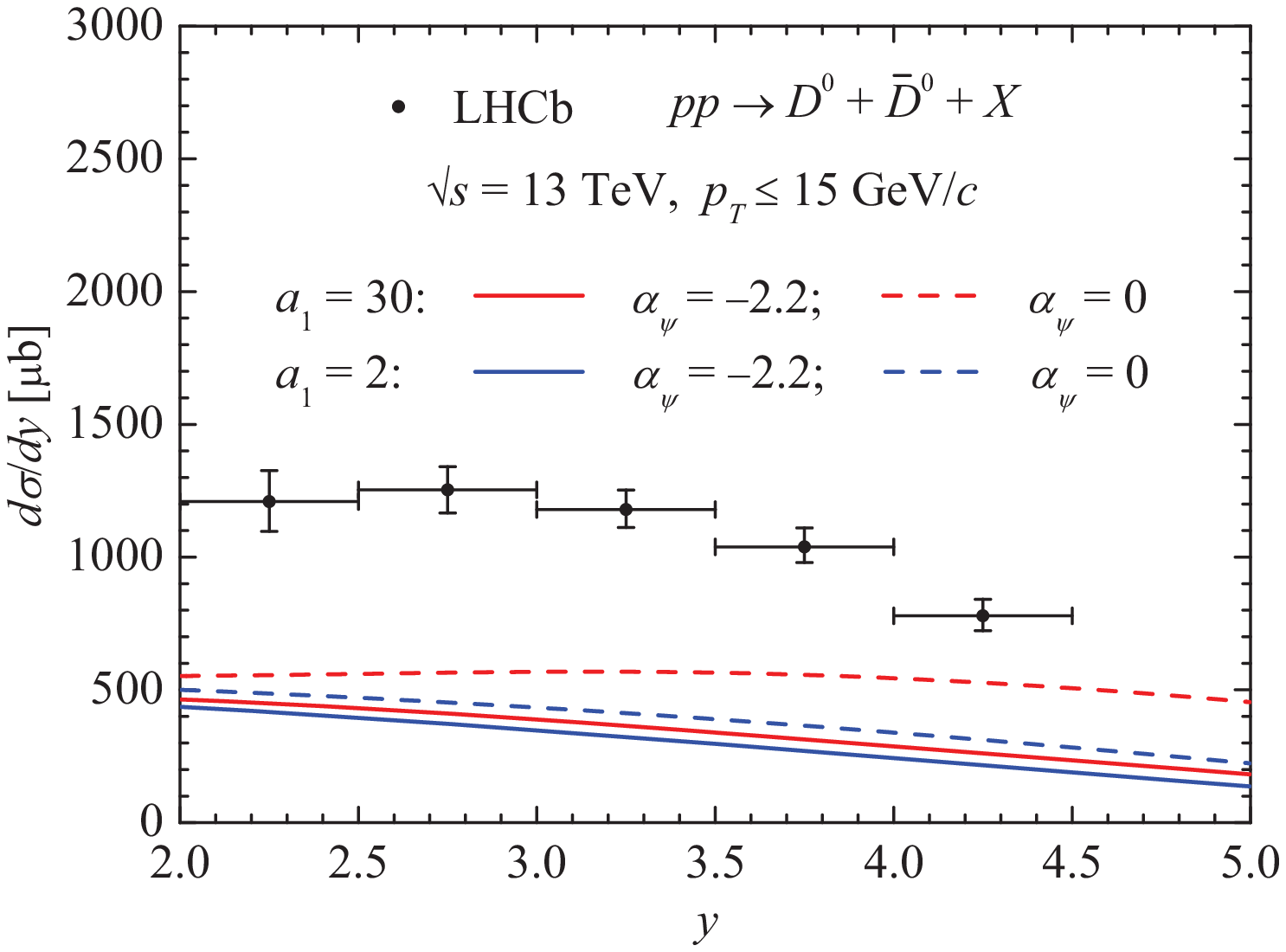}
	\caption{\label{fig:11}Differential cross sections of charged $D$ mesons (left) and neutral $D$ mesons (right) in $pp$ collisions at $\sqrt{s}=13$ TeV. Data are from LHCb experiment~\cite{LHCb13TeV}. The same notation for lines as in Fig.~\ref{fig:9}.}
\end{figure*}
As can be seen from Figs.~\ref{fig:2},~\ref{fig:3}, QGSM with the intercept $\alpha_{\psi}(0)=-2.2$  better describes the experimental data for differential cross sections. 
Figure~\ref{fig:3} presents the comparison of the experimental data ($pp$ collisions, 400 GeV)~\cite{NA27} with calculations of differential cross sections for each sort of $D$ mesons production ($D^{+}, D^{-}, D^{0}, \bar{D}^0$).  The cross sections of $D^{+}$ and $D^{0}$ weakly depend on the parameter~$a_{1}$, while cross sections of $D^{-}$ and $\bar{D}^0$ mesons  calculated for $a_{1}=2$ are smaller (e.g. for $\alpha_{\psi}(0)=-2.2$ by a factor~$2-10$) as compared to the case of $a_{1}=30$. The intercept $\alpha_{\psi}(0)=-2.2$ noticeably better describes  the experimental data  for $D^{-}$ and $\bar{D}^0$,   whereas values $\alpha_{\psi}(0)=0$, $a_{1}=2$  lead to better description  of $D^{0}$ mesons. In spite of data spread for $D^{+}$, the intercept $\alpha_{\psi}(0)=-2.2$  seems preferable.      

The factor $a_{1}$ amplifies the contribution of the leading fragmentation ($D^{-}/\bar{D}^0$) with participation of the valence quarks. 
Production of $D^{-}$ and $\bar{D}^0$ in $pp$~interactions has a higher probability because these mesons contain the valence quarks of colliding protons. The contribution of the leading fragmentation functions dominates, and $x$-distribution of $D^{-}$ and $\bar{D}^0$ is harder in comparison with $D^{+}$ and $D^{0}$. The influence of the parameter~$a_{1}$ on the cross section of all $D$~mesons production is also noticeable (Fig.~\ref{fig:2}). 
\par
 The calculations of cross sections of charmed meson production in $\pi^{-}p$ collisions are compared to experimental data in Figs.~\ref{fig:7}--\ref{fig:6}. 
Figure~\ref{fig:7} shows the differential cross sections of neutral $D^{0}/\bar{D}^{0}$ mesons production in $\pi^{-}p$ collisions at energy 500 GeV in comparison to the data of the experiment E791~\cite{E791}. The calculation  for values $\alpha_{\psi}(0)=-2.2$ and $a_{1}=2$ agrees with the measurement data in the small $x$ range, and prediction obtained for $\alpha_{\psi}(0)=-2.2$ and $a_{1}=30$ better describes the experimental data in the fragmentation region.
The differential cross sections of $D/\bar{D}$ mesons computed at energy 360 GeV in comparison to the  data of experiments  WA92 (350 GeV)~\cite{WA92} and NA27 (360 GeV)~\cite{NA27_pip} are presented in Fig.~\ref{fig:4}. 
In spite of small differences of the beam energies, data of the two experiments differ at small-$x$ and at $x>0.5$. The calculation for $\alpha_{\psi}(0)=-2.2$ and $a_{1}=2$ describes well most of experimental points. 
\par
The differential cross sections for each type of $D$ mesons in comparison to measurements (WA92) are shown in Fig.~\ref{fig:5}. Calculations with $\alpha_{\psi}(0)=-2.2$ have good agreement with the data on the neutral $D$ meson production. On the other hand the intercept $\alpha_{\psi}(0)=-2.2$ gives overestimated cross sections in case of charged $D$ meson production at $x<0.5$. 
As one can see from Figs.~\ref{fig:4}--\ref{fig:6}, there is similar dependence on the parameter $a_{1}$ of $D$ meson cross sections in $pp$ collisions (see Figs.~\ref{fig:2},~\ref{fig:3}): $x$-distributions of leading particles are more sensitive to $a_{1}$. The only difference is that $D^{-}$ and $D^{0}$ produced in $\pi^{-}p$ interactions are leading particles unlike  $D^{+}$ and $\bar{D}^{0}$ by virtue of different quark composition of the colliding particles.
\par
The calculations of the leading and nonleading differential cross sections of $D$ mesons production in $\pi^{-}p$ collisions at energy 360 GeV (LF) are shown in Fig.~\ref{fig:6} along with experimental data of NA27~\cite{NA27_pip}. %
\par
Comparison of the cross section of ${\rm\Lambda}^{+}_{c}$~baryon production in $pp$ collisions with experimental data is shown in Fig.~\ref{fig:8}. 
The differential cross section was calculated at $\sqrt{s}=62$~GeV (left panel) the experimental data were obtained for energies  $\sqrt{s}=62$~GeV~\cite{Lambda_Bari} and $63$~GeV~\cite{Lambda_Chauvat}.
There is appreciable difference of the cross section  measurements in  these two experiments. 
The calculation with the parameter~$\alpha_{\psi}(0)=-2.2$ agrees with the later experiment.

The right panel of Fig.~\ref{fig:8} shows the total cross section of ${\rm\Lambda}^{+}_{c}$~production as a function of center-of-mass energy. The experimental points are taken from Ref.~\cite{Lambda_Bari}, the calculation was made for the same parameter sets. The large spread of the total cross section data prevents from making definite choice of the intercept $\alpha_{\psi}(0)$. 
\par
In Figs.~\ref{fig:9}--\ref{fig:11}, we show the differential cross sections of $D$ mesons 
production in $pp$ collisions as a function of the rapidity ($y$) in comparison to 
LHCb measurements. Experimental data were obtained at energies $\sqrt{s}=5$ TeV~\cite{LHCb5TeV}, 7 TeV~\cite{LHCb7TeV} and 13 TeV~\cite{LHCb13TeV} for rapidity range $2\leqslant y \leqslant4.5$ that corresponds to $x \lesssim 10^{-3}-10^{-2}$. 
The points plotted in Figs.~\ref{fig:9}--\ref{fig:11} were obtained from original experimental data by  
summing them over transverse momentum bins (for each bin in $y$). 
\par
There is expected the problem in describing the experimental data on $D$ mesons   
in the small $x$ range. In the QGSM version under consideration, the inclusive spectra of charmed particles production are averaged over transverse momentum, while the LHCb data were obtained for the transverse momentum interval $p_{\perp}\leqslant 15$ GeV/c at the rapidity values $2.0-4.5$. However the calculation with $\alpha_{\psi}(0)=-2.2$  describes satisfactorily the experimental measurements on the production $D^{\pm}$  mesons (unlike $D^{0}/{\bar{D}}^{0}$) at energies $\sqrt{s}=5$ TeV and $7$ TeV  (Figs.~\ref{fig:9},~\ref{fig:10}).
\begin{figure*} 
	\centering	
		\includegraphics[trim=0cm 0cm 0cm 0cm, width=0.46\textwidth]{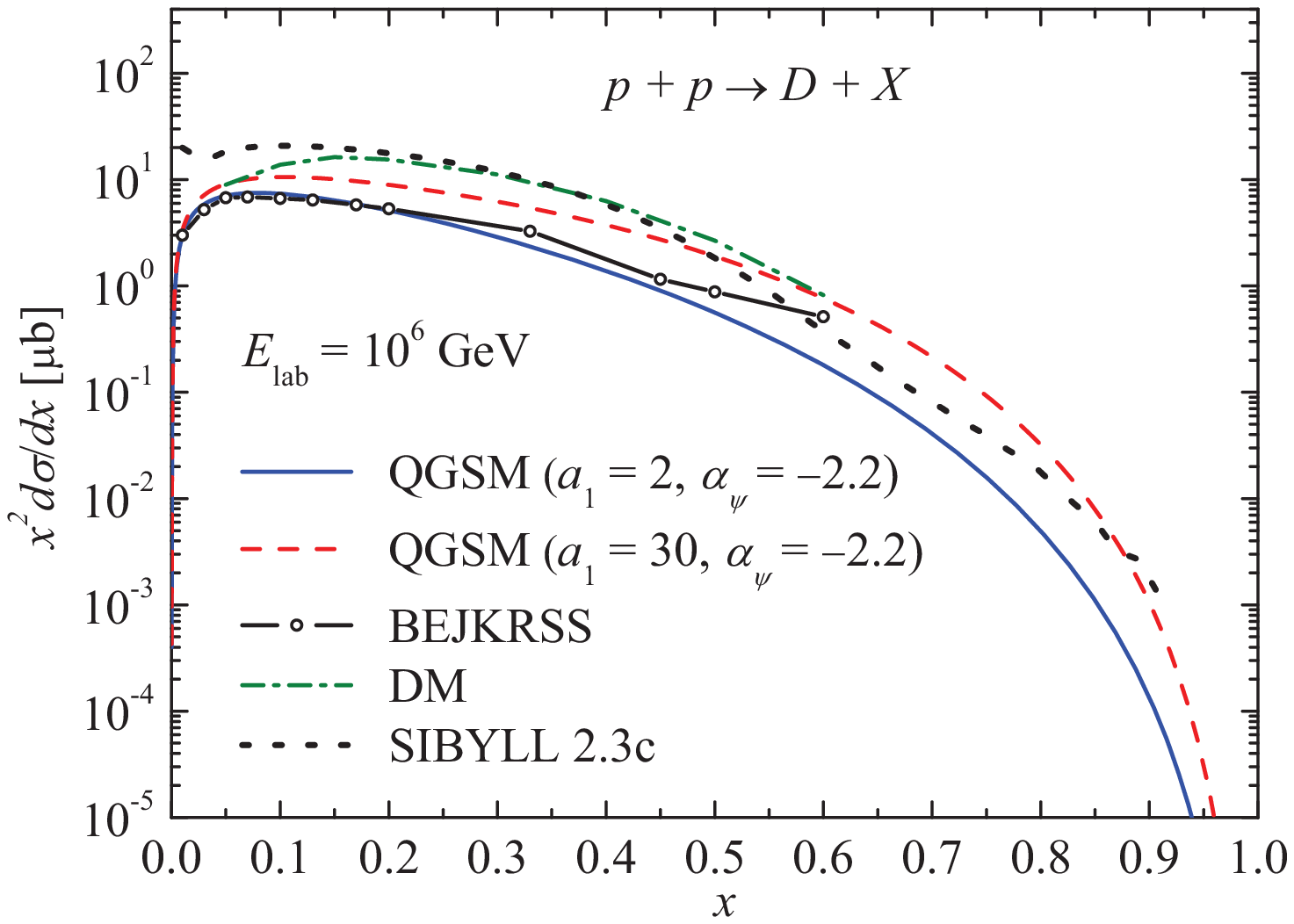}
	\hspace{3mm}
		\includegraphics[trim=0cm 0cm 0cm 0cm, width=0.46\textwidth]{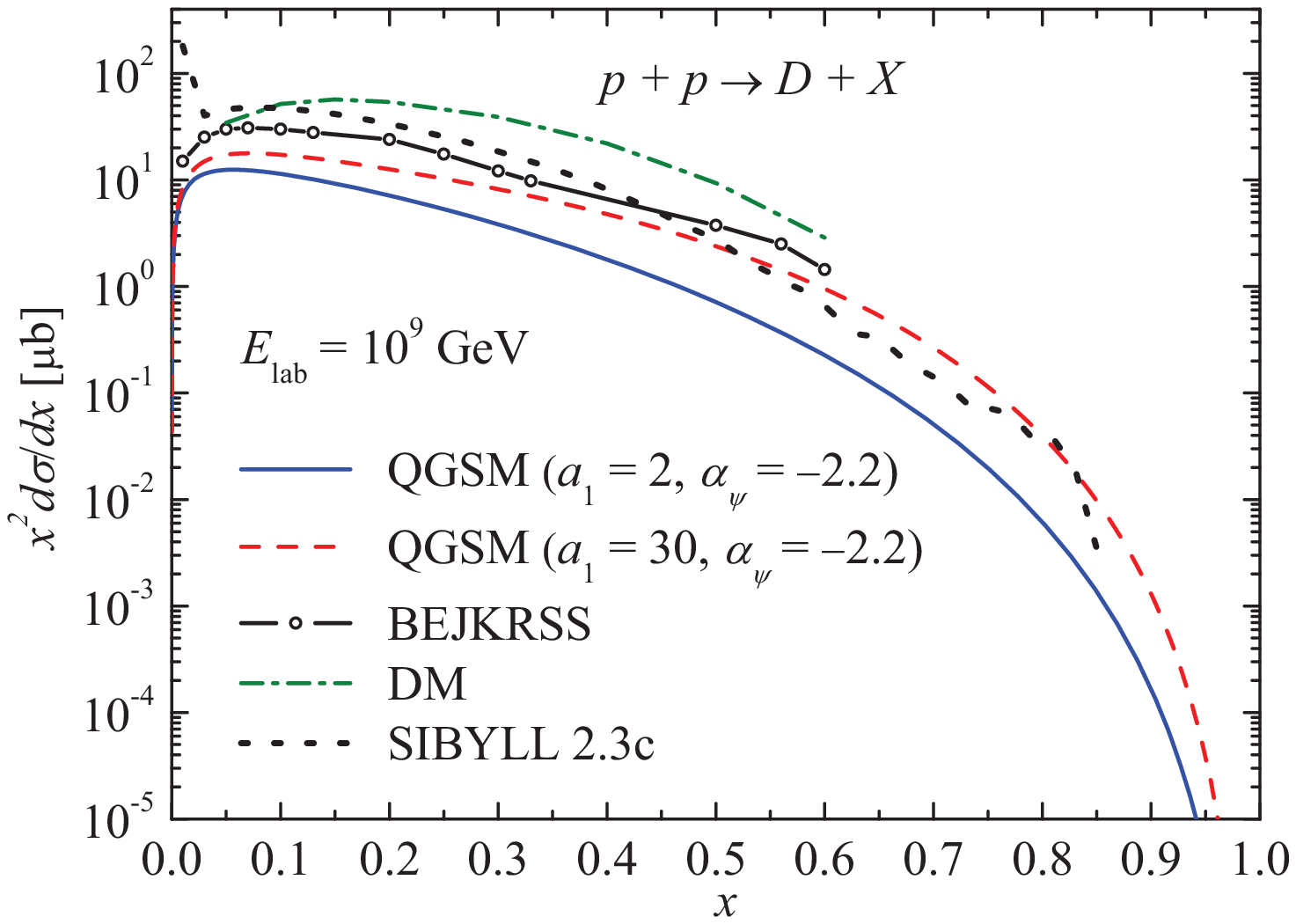}
	\hspace{3mm}
		\includegraphics[trim=0cm 0cm 0cm 0cm, width=0.46\textwidth]{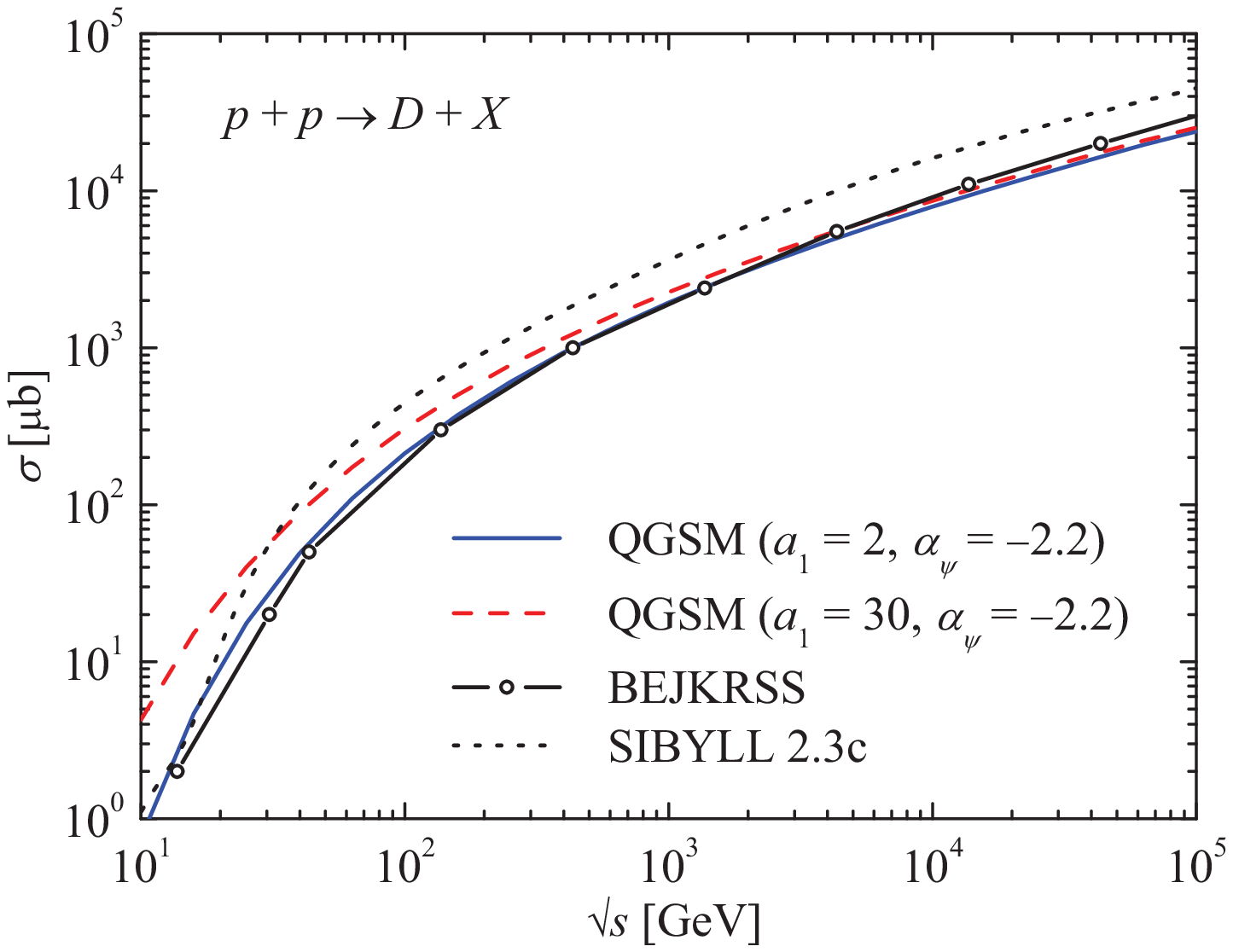}
	\caption{\label{fig:13} Model predictions for the differential (top) and total (bottom) cross sections of $D/\bar{D}$~mesons production in $pp$ collisions: QGSM with $\alpha_{\psi}(0)=-2.2$ for $a_{1}=2$ (solid line) and $a_{1}=30$ (dash), NLO pQCD (BEJKRSS)~\cite{BEJKRSS} (line with circles), dipole model (DM)~\cite{ERS} (dash-dotted) and 
	SIBYLL 2.3c~\cite{sib-2.3c_18} (dotted).} 
\end{figure*}
The experimental data on neutral $D^{0}/\bar{D}^{0}$ and charged $D^{+}/D^{-}$ mesons differ by 
factor $~2-3$ (at fixed energy), while QGSM predicts close values of $d\sigma/dy$. That is the model describes the cross sections of charged mesons much better than $D^{0}/\bar{D}^{0}$. One possible explanation of this discrepancy is that $D^{0}/\bar{D}^{0}$ events contain  a mixture of vector mesons $D^{*0}/\bar{D}^{*0}$, decays of which might contribute to pseudoscalar $D^{0}$ mesons.


 
It is possible also that in case of the small-$x$ events  and  large transverse momentum at high energy we encounter with the problem of ``enhanced'' diagrams, relating to interactions  between  pomerons, which are neglected in the  quasieikonal approximation. Account of enhanced diagrams leads to $x$-distributions, rising  as $1/x$ at small $x$ \cite{Kaidalov03_YaF}. Perhaps it will take a significant revision of the  QGSM parameters and calculation technique for this region of kinematics.  

Note however that small-$x$ region ($10^{-4}-10^{-3}$) gives minor contribution to the atmospheric neutrino flux because of the dominating peripheric  processes in the cosmic-ray induced hadronic cascade: the small-$x$ values are suppressed under the integral by a factor $x^{\gamma}$,  where $\gamma$ is the spectral index of cosmic ray protons ($\gamma\approx~ 1.7-2.0$).

\section{QGSM in comparison with different charm production models}
\label{sec:charmmodels}
Before comparing of the prompt atmospheric neutrino fluxes predictions it would be useful to confront cross sections of charmed particles production of different models. The comparison of the differential cross sections of charmed mesons production in $pp$ collisions for proton energies in the laboratory frame ($10^{3}$ and $10^{6}$ TeV) is shown in the top panel of Fig.~\ref{fig:13}:  QGSM (solid and dashed lines), 
 SIBYLL 2.3c~\cite{sib-2.3c_18} (dotted line), perturbative QCD model (BEJKRSS)~\cite{BEJKRSS} (the line with symbols), and  the dipole model (DM)~\cite{ERS} (dash-dotted). The pQCD calculation is rather close to the present work results obtained with parameters $\alpha_{\psi}(0)=-2.2$, $a_{1}=2$ ($10^{3}$ TeV) and $a_{1}=30$ ($10^{6}$ TeV). Our calculation lies below the DM result for most of the $x$ range, that should lead to the lowered prompt neutrino flux as compared with the result of Enberg et al.~\cite{ERS}.
\par %
The total cross sections of $D$ mesons production in $pp$ collisions as a function of center-of-mass energy are shown in the bottom panel of Fig.~\ref{fig:13} for QGSM, SIBYLL 2.3c~\cite{sib-2.3c_18} and pQCD model (BEJKRSS)~\cite{BEJKRSS}. Predictions of the QGSM (for $a_{1}=2$) and pQCD model are almost the same in a wide energy range, with the exception of energies $\sqrt{s}<100$ GeV (calculation for $a_{1}=30$ gives large cross sections at $\sqrt{s}<10^{3}$ GeV).                  
 
To calculate the prompt neutrino flux one needs know the cross sections of charmed particles production in collisions of hadrons with atmospheric nuclei. The cross sections are recalculated for a nuclear target with average mass number $A$ according to the formula $d\sigma_{pA}/dx=A^{\alpha}d\sigma_{pp}/dx$ (for the air we take $A=14.5$). The index $\alpha$ depends on $x$: 
$\alpha\approx1$ at $x\rightarrow0$ and monotonically decreases with rise of $x$ ($\alpha\approx0.5$ at $x\rightarrow1$)~\cite{KaidalovPiskunova_charm}. In~\cite{BNSZ} the prompt neutrino flux has been calculated for constant $\alpha\approx0.72$ (authors argued that uncertainty due this choice does not exceed 15\%), and we use the same value.
\section{Energy spectra of the prompt atmospheric neutrinos}
\label{sec:fluxes}
\begin{figure*} 
	\centering
		\includegraphics[trim=0cm 0cm 0cm 0cm, width=0.46\textwidth]{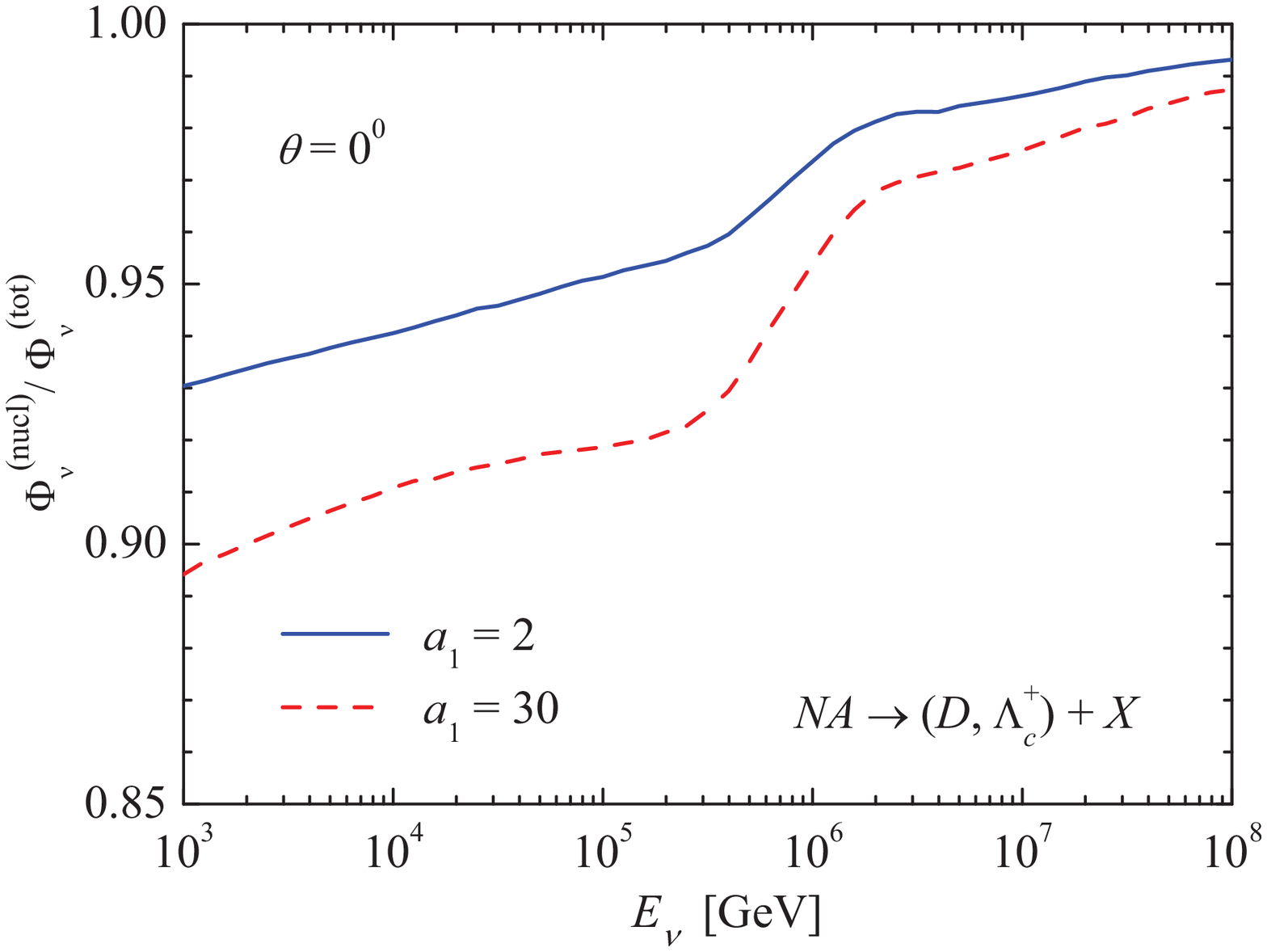}
	\hspace{3mm}
		\includegraphics[trim=0cm 0cm 0cm 0cm, width=0.46\textwidth]{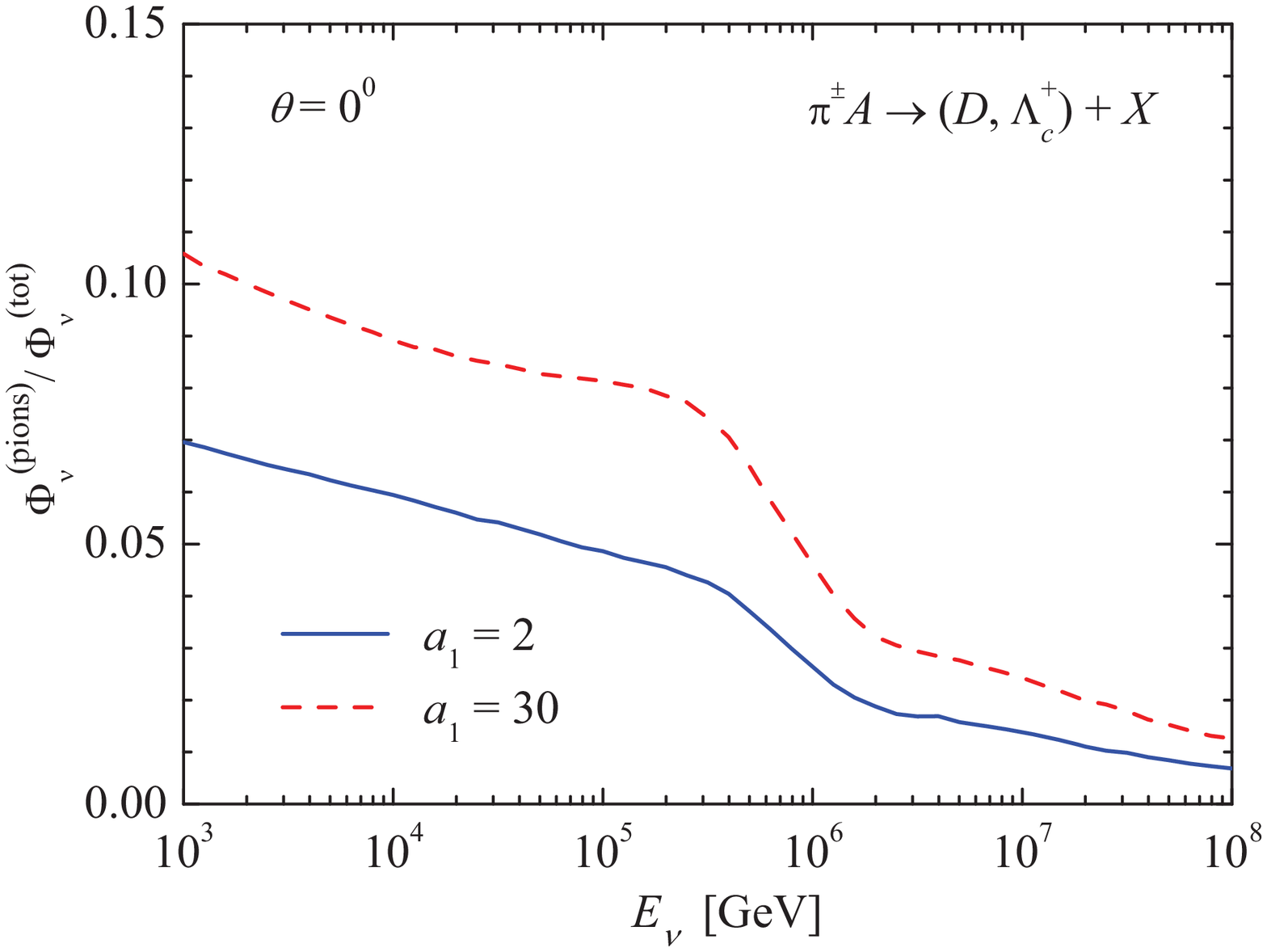}
	\caption{\label{fig:14}Partial contributions of $NA$  (left) and $\pi A$ collisions (right) to the  prompt muon neutrinos calculated for NSU spectrum with $\alpha_{\psi}(0)=-2.2$ for $a_{1}=2$  (solid line) and $a_{1}=30$ (dash).}
\end{figure*}
\begin{figure*} 
	\centering
		\includegraphics[trim=0cm 0cm 0cm 0cm, width=0.46\textwidth]{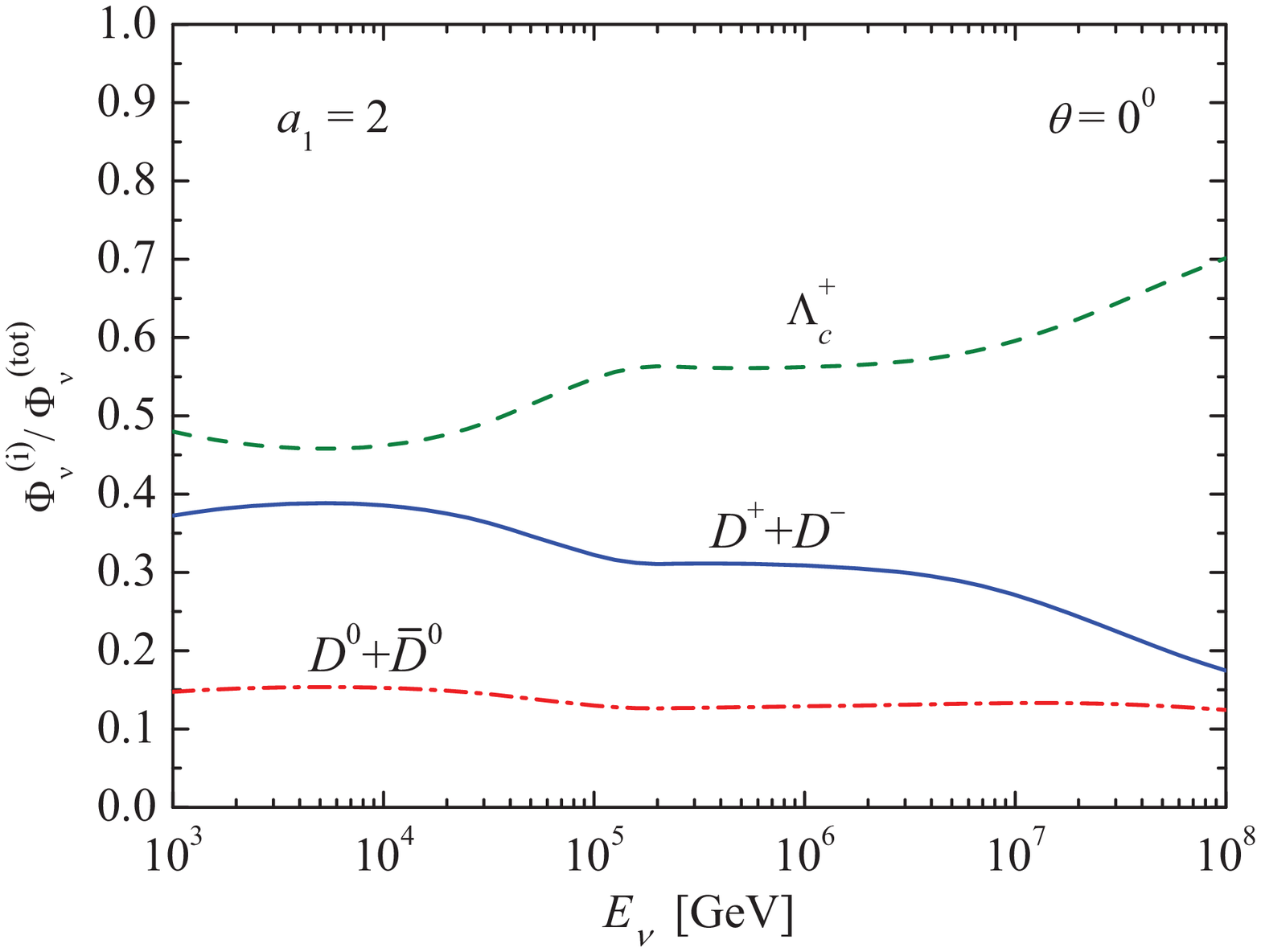}
	\hspace{3mm}
		\includegraphics[trim=0cm 0cm 0cm 0cm, width=0.46\textwidth]{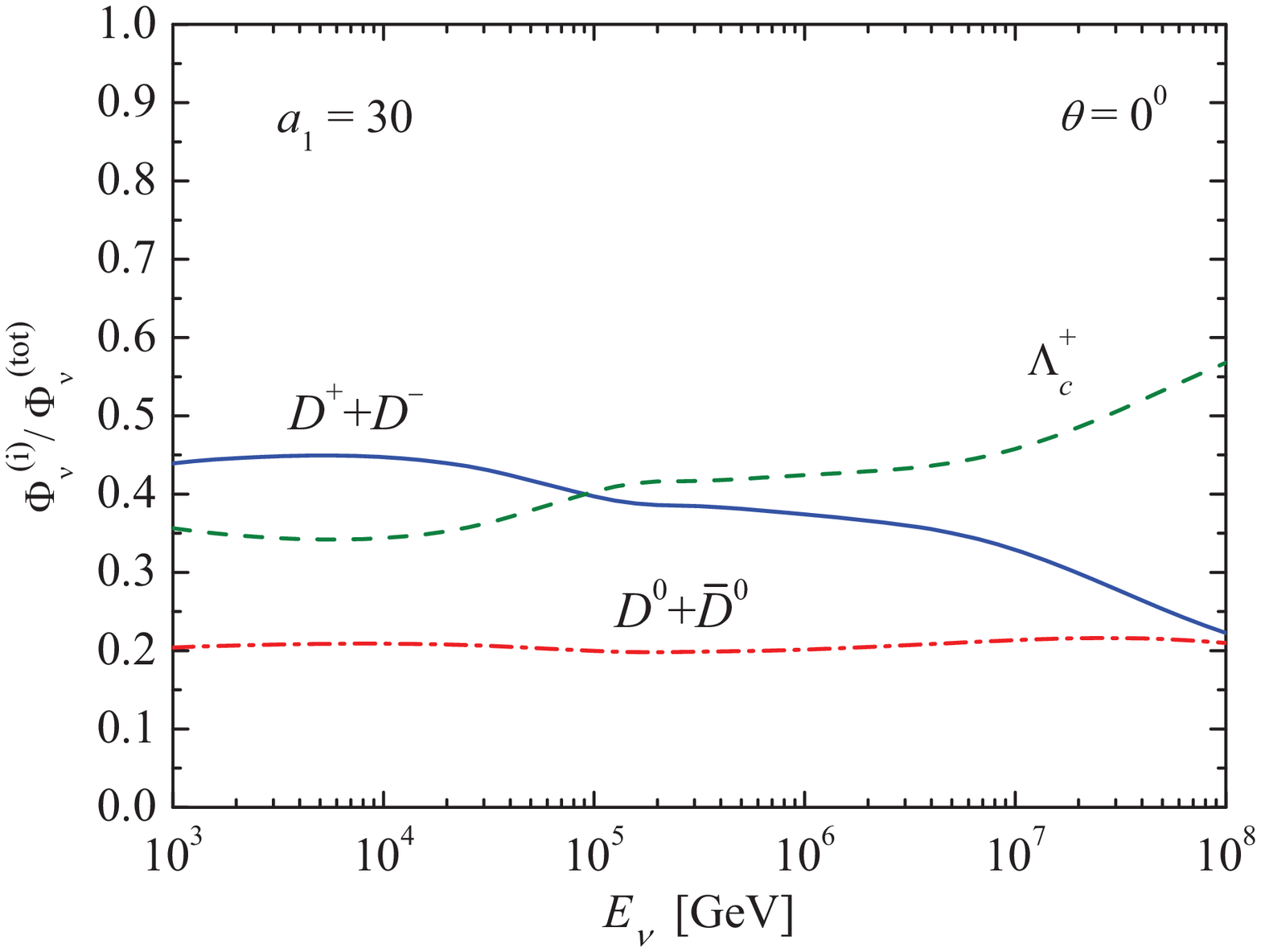}
	\caption{\label{fig:15}Partial contributions to the prompt muon neutrino flux calculated with QGSM for NSU spectrum with $\alpha_{\psi}(0)=-2.2$ for $a_{1}=2$ (left) and $a_{1}=30$ (right).} 
\end{figure*}
In the present work, the calculation of prompt neutrino fluxes is performed with the method \cite{BNSZ,Vall} for QGSM for parameterization of cosmic ray spectrum by Nikolsky, Stamenov, Ushev (NSU)~\cite{NSU}, the toy model by Thunman, Ingelman, Gondolo (TIG)~\cite{TIG}, and the recent model for cosmic ray spectrum by Hillas and Gaisser (H3a)~\cite{H3a}. The NSU spectrum which takes into account an elemental composition of primary cosmic rays was chosen in order to compare new result with the old one~\cite{BNSZ,nss98}. We use also the toy spectrum TIG (else called the broken power law, BPL) only in order to compare results of different calculations, including the dipole model prediction~\cite{ERS} which was used by IceCube as a benchmark model. The three-component model with mixed extragalactic population, H3a was chosen to compare the QGSM calculation with  SIBYLL 2.3c~\cite{sib-2.3c_18}, the NLO pQCD predictions BEJKRSS~\cite{BEJKRSS} and GRRST \cite{grrst},  as well as with the IceCube experiment limitation on the prompt neutrino flux~\cite{IceCube16}.   %
\par  
Nucleon-nuclear interactions give the main contribution to the prompt neutrino flux, while reactions $\pi^{\pm}A$ add less than $5$-$10$\% to the prompt neutrino flux in the energy range $10-10^{5}$ TeV. The Fig.~\ref{fig:14} shows relative contributions of $NA$ and $\pi A$ interactions to the prompt fluxes calculated for NSU spectrum and two values of the free parameter of the quark-gluon string model: $a_{1}=2$ (solid line) and $a_{1}=30$ (dashed one). Contributions 
of $D$ mesons and ${\rm\Lambda}^{+}_{c}$~baryons are presented in Fig.~\ref{fig:15}. 
\par
\begin{figure} 
	\centering
		\includegraphics[trim=0cm 0cm 0cm 0cm, width=0.46\textwidth]{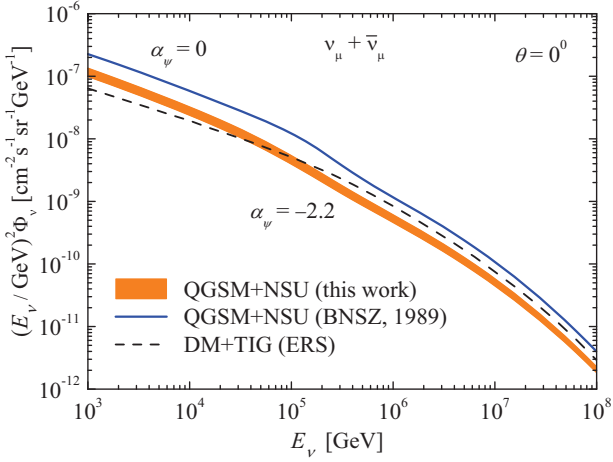}
	\caption{\label{fig:16} Vertical flux of prompt atmospheric neutrinos calculated with QGSM for the NSU cosmic ray spectrum.  Shaded band: this work with $\alpha_{\psi}(0)=-2.2$. The band width corresponds to the  variation of $a_1$ from $2$ (lower bound) to $30$ (upper bound). The solid line presents the result from Refs.~\cite{BNSZ,nss98} obtained with $\alpha_{\psi}(0)=0$. Dash line plots the dipole model calculation~\cite{ERS} for the TIG spectrum.}
\end{figure}

 
Figure~\ref{fig:16} shows the calculation of the prompt atmospheric neutrinos flux (scaled by $E^{2}_{\nu}$): the band represents this work calculation for the NSU spectrum and the QGSM with the intercept~$\alpha_{\psi}(0)=-2.2$. The band shows uncertainty due to change of the parameter~$a_{1}$ which ensures unified behavior of the leading fragmentation functions at $x\rightarrow 0$ and  $x\rightarrow 1$. Extreme values of~$a_{1}$ lead to change of the neutrino flux by a factor~$1.4$: $a_{1}=2$ corresponds to lower bound, and $a_{1}=30$ to upper one.
However, influence of the intercept of Regge trajectory~$\alpha_{\psi}(0)$ appears to be more substantial: the replacement of~$\alpha_{\psi}(0)=0$ by~$\alpha_{\psi}(0)=-2.2$ reduces the flux by a factor~3 as compared to the result~\cite{BNSZ} (solid line) obtained for the same scheme (QGSM+NSU) with intercept~$\alpha_{\psi}(0)=0$. The dipole model result~\cite{ERS} for the TIG cosmic ray spectrum is also shown in Fig.~\ref{fig:16} (dashed line).

The QGSM flux~\cite{BNSZ} performed for $\alpha_{\psi}(0)=0$ and NSU spectrum was considered by IceCube collaborators as too optimistic prediction~\cite{IC14,IC15}. At the energies $E_{\nu}>10^6$~GeV it exceeds the ERS result~\cite{ERS} by about 30\%, however part of this excess is related to the difference of the cosmic ray spectra used. 
\par
\begin{figure} 
	\centering
		\includegraphics[trim=0cm 0cm 0cm  0cm, width=0.47\textwidth]{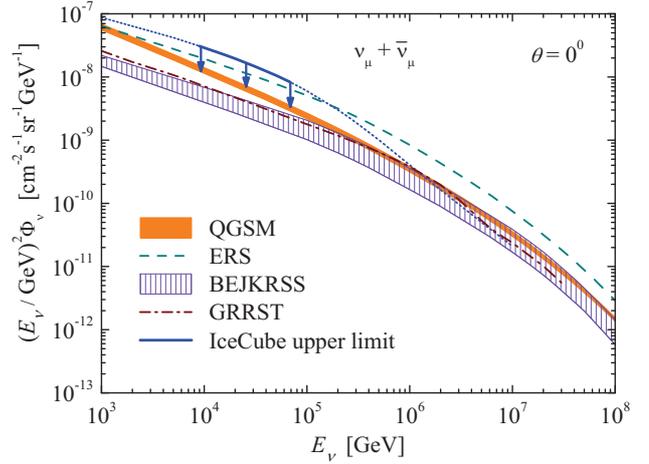}	
	\caption{\label{fig:17} Comparison of the prompt neutrino flux  calculated with use of the models: QGSM with $\alpha_{\psi}(0)=-2.2$ (shaded band), the color dipole model (ERS)~\cite{ERS} (dash line), NLO pQCD models, BEJKRSS~\cite{BEJKRSS} (hatched area) and  GRRST \cite{grrst} (dash-dotted line). All calculations are performed for the TIG cosmic ray spectrum.  Solid line denotes the IceCube upper limit and the dotted one its extrapolation~\cite{IceCube16} (see text for more details).} %
\end{figure}

An influence of charm production models on neutrino fluxes is seen in Fig.~\ref{fig:17}. All results are obtained for the same cosmic ray spectrum TIG.  
 In the energy range beyond 1 PeV, where atmospheric neutrinos from charmed particles dominate, QGSM (shaded band) leads to appreciably lower flux as compared to the dipole model result~\cite{ERS} (dashed line). The predictions of the pQCD models, BEJKRSS~\cite{BEJKRSS} and GRRST \cite{grrst} are compatible in the whole energy range. The QGSM flux at energies above 200 TeV is close  to upper bound of the BEJKRSS band.
\begin{figure} 
	\centering
		\includegraphics[trim=0cm 0cm 0cm 0cm, width=0.46\textwidth]{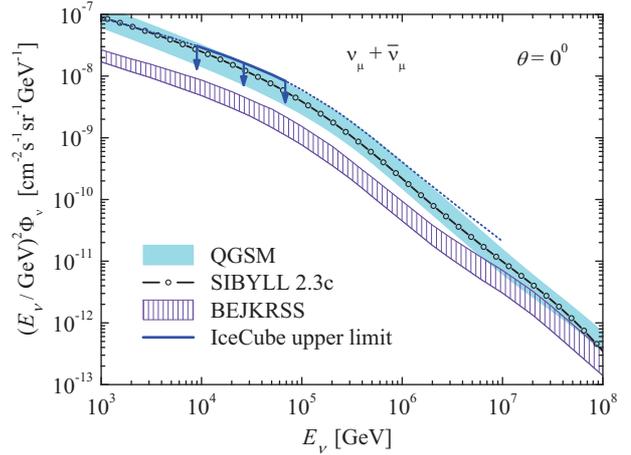}	
	\caption{\label{fig:18add} Uncertainties of the prompt neutrino flux calculations.
	The blue band corresponding to QGSM calculations shows the effect of varying crucial
parameter:  the intercept $\alpha_{\psi}(0)=0$ (upper bound of the band) and $\alpha_{\psi}(0)=-2.2$ (lower bound). The hatched area presents the scale of uncertainties of the flux calculation with the NLO pQCD model BEJKRSS~\cite{BEJKRSS}. The SIBYLL 2.3c~prediction \cite{sib-2.3c_18} is compatible with QGSM (the line inside band). All calculations are performed for the H3a cosmic ray spectrum. The solid line with arrows represents the IceCube limitation and dotted line labels its extrapolation~\cite{IceCube16}.} %
\end{figure} 
\begin{figure} 
	\centering
		\includegraphics[trim=0cm 0cm 0cm 0cm, width=0.47\textwidth]{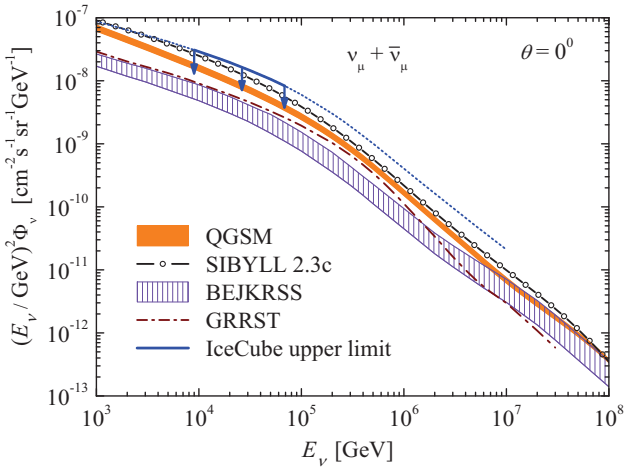}
	\caption{\label{fig:18} The comparison of the prompt neutrino flux calculated for the H3a cosmic ray spectrum~\cite{H3a}  with use of different models: QGSM ($\alpha_{\psi}(0)=-2.2$), SIBYLL 2.3c~\cite{sib-2.3c_18}, the pQCD models  BEJKRSS~\cite{BEJKRSS} and GRRST~\cite{grrst}.
	Solid line is the IceCube constraint, dotted line plots the limit extrapolation~\cite{IceCube16}.} 
\end{figure}
\par  
The uncertainty of the QGSM neutrino flux prediction due to variation of the intercept $\alpha_{\psi}(0)$ is shown in Fig.~\ref{fig:18add}. Uncertainty band is computed for $\alpha_{\psi}(0)=-2.2$ with $a^{D}=7.0\cdot10^{-4}$ (lower bound)  and $\alpha_{\psi}(0)=0$ with $a^{D}=10^{-3}$ (upper bound). The band calculation is performed for H3a cosmic ray spectrum  with use the value  $a_{1}=2$.
The bulk of prompt neutrino flux uncertainties is induced by intercept of the charm Regge trajectory (as Fig.\ref{fig:18add} shows). Rest parameters, $a^{D}$,  $a_{1}$ slightly disturb bounds of the uncertainty span. In fact, the uncertainty due to the  $\alpha_{\psi}(0)$-variation absorbs that for the $a_{1}$-variation.

 $\alpha_{\psi}(0)=-2.2$ seems to be preferable from the viewpoint of the most experimental measurements on charm production. However, $\alpha_{\psi}(0)=-2.2$ leads to underestimation of the neutral $D$ meson production at high energy. On the other hand  $\alpha_{\psi}(0)=0$ predicts harder charm spectra at all $x$. The largest differences in $d\sigma/dx$ appear at $x\rightarrow 1$, where the charm spectra calculated for $\alpha_{\psi}(0)=0$ are larger by several orders of magnitude in comparison with case of $\alpha_{\psi}(0)=-2.2$. The $x\sim 1$  is the region of peripheric processes which dominate in the atmospheric hadronic cascade and  atmospheric neutrino fluxes turn out to be sensitive to the intercept value: the replacement of $\alpha_{\psi}(0)=-2.2$  by $\alpha_{\psi}(0)=0$ increases flux by a factor $2-3$. 
 
Relative to the BEJKRSS result, the QGSM gives larger flux at low and middle energies. However, our calculation comes near to the  BEJKRSS one with energy growth, and the QGSM lower bound is compatible with upper bound of the pQCD prediction at $E_{\nu}>10$ PeV. On the other hand, our uncertainty band completely covers the SIBYLL 2.3c flux in the whole energy range.

Figure~\ref{fig:18} presents our preferred result: the  prompt neutrino spectrum calculated with QGSM for the  magnitudes of parameters $\alpha_{\psi}(0)=-2.2$,  $a^D=7\cdot 10^{-4}$.  
The brown band marks the QGSM uncertainty related to the parameter $a_1$ which rises from $2$ (lower bound) to $30$ (upper bound). 
Also shown here are the SIBYLL 2.3c~\cite{sib-2.3c_18} calculation  and the NLO  pQCD ones (BEJKRSS~\cite{BEJKRSS} and GRRST~\cite{grrst} models). All these spectra were obtained for H3a cosmic ray spectrum. The QGSM flux for $a_{1}=30$ and $\alpha_{\psi}(0)=-2.2$ is close to the  SIBYLL 2.3c calculation beyond $100$ TeV and  greater than the BEJKRSS flux by a factor $\sim 2-4$ at $10^5-10^6$ GeV. With the energy rise these differences decrease, and our prediction at $E_{\nu}\gtrsim 10$ PeV is close to the BEJKRSS flux. 

In the IceCube analysis~\cite{IceCube16}, upper limit on the prompt atmospheric neutrino flux was obtained using high statistics collected over six years. The prompt neutrino flux was constrained using the color dipole model~\cite{ERS} corrected for the cosmic ray spectrum parametrization H3p~\cite{H3a}.   
The solid line with arrows in Figs.~\ref{fig:17},\ref{fig:18add},\ref{fig:18} shows the constraint specified by IceCube for the sensitive energy range $9-69$ TeV. 
The  extrapolation beyond $69$ TeV was made by IceCube using again the prompt neutrino spectrum~\cite{ERS}: the original result was also brought into accord with the cosmic-ray model H3p~\cite{H3a}. 
The dotted line in Figs.~\ref{fig:18add},\ref{fig:18} just shows the dipole model behavior beyond of the sensitivity range with the best-fit parameters describing the systematic uncertainties of the IceCube analysis, e.g., the optical efficiency of the telescope, the Antarctic ice properties, the uncertainties of the cosmic ray spectrum. 
 
Our result is evidently compatible with the IceCube limitation, the same is true for the rest models under discussion: SIBYLL 2.3c~\cite{sib-2.3c_18}, the NLO pQCD models  BEJKRSS \cite{BEJKRSS} and GRRST~\cite{grrst}.

\section{Conclusion}
\label{sec:conclusion}

The new calculation of the atmospheric neutrino flux from decays 
of the charmed particles is performed with updated version of the quark-gluon string model. 
The QGSM parameters~$\alpha_{\psi}(0)$ and~$a_{1}$ are examined by comparison of calculated cross sections for the charmed meson production with the data of measurements obtained at LHC  and  in other experiments. 
Though the data of the LHCb experiment do not allow unique choice of the varying parameters magnitudes, we consider the intercept  $\alpha_{\psi}(0)=-2.2$ as a preferred value against $\alpha_{\psi}(0)=0$.   Note, that  $\alpha_{\psi}(0)=-2.2$ appears proper magnitude, because it is in accordance also with observable (and natural) pattern: heavier quarks have lower intercept of the Regge trajectory. 

At high energy the differential cross sections of charm production are more sensitive (as compared with the total cross section) to the change of the parameter $a_{1}$, which brings the neutrino flux deviation  about $(20-40)\%$ for extreme values of $a_{1}$. The analysis shows that intercept of Regge 
trajectory~$\alpha_{\psi}(0)$ causes more noticeable effect on the charm production and 
therefore on the prompt atmospheric neutrino flux. Updated version of QGSM with $\alpha_{\psi}(0)=-2.2$ 
leads to decrease of the prompt neutrino flux by a factor~$\sim 2-3$ as compared to the 
former QGSM prediction \cite{BNSZ,nss98} obtained with $\alpha_{\psi}(0)=0$. 
\par                             
In the energy range beyond 1 PeV, where atmospheric neutrinos from the decay of charmed particles dominate, as it is expected, the new QGSM flux is significantly lower in comparison with the color dipole model~\cite{ERS}.  The QGSM flux obtained for intercept~$\alpha_{\psi}(0)=-2.2$ and H3a cosmic ray spectrum is compatible with the NLO pQCD predictions at $E_{\nu}>10$ PeV, and upper bound of our calculation does not differ practically from the SIBYLL 2.3c result. The updated QGSM calculation of the prompt atmospheric neutrino flux is consistent with the IceCube limitation.

The performed calculations confirm viability of QGSM as appropriate phenomenological model of the high-energy hadronic interactions which allows account of effects beyond the pQCD.
The updated version of QGSM is the suitable approach to provide reasonable prediction of the atmospheric prompt neutrino flux.   
Undoubtedly, the current version QGSM must be developed taking into account the large-mass diffraction  dissociation and resonances production. Also all  parameters of the model need in comprehensive revision and the thorough fitting by use of recent and future results of high-energy hadronic experiments.

\section*{Acknowledgements}
We are grateful to A. Fedynitch and J. Talbert for kindly providing us with tables of the prompt neutrino flux calculations.

\end{document}